\RequirePackage[2024-06-01]{latexrelease}
\documentclass[trackchanges, twocolumn]{aastex701}
\usepackage{xspace}
\usepackage{array}
\usepackage{threeparttable}
\usepackage{bm}
\usepackage{amsmath}
\usepackage{color,soul}

\newcommand{\aestra}{\textit{\AE{}stra}\xspace}
\newcommand{\metersec}{m\,s$^{-1}$}

\defcitealias{2013A&A...558A..33A}{Astropy Collaboration et al. 2013}
\defcitealias{2018AJ....156..123A}{Astropy Collaboration et al. 2018}
\defcitealias{2022ApJ...935..167A}{Astropy Collaboration et al. 2022}
\defcitealias{ChatGPT2026}{Open AI 2026}

\begin{document}

\title{AESTRA II: Generative Spectral Modeling of the Sun as a Star for Precise Radial Velocities}

\author[0000-0002-1001-1235]{Yan Liang}
\affiliation{Department of Astrophysical Sciences, Princeton University,
Princeton, NJ 08544, USA}
\affiliation{Department of Astronomy, Yale University,
New Haven, CT 06511, USA}
\email[show]{yan.liang@yale.edu}
\author[0000-0002-4265-047X]{Joshua N.\ Winn}
\affiliation{Department of Astrophysical Sciences, Princeton University,
Princeton, NJ 08544, USA}
\email{jnwinn@princeton.edu}
\author[0000-0002-8873-5065]{Peter Melchior}
\affiliation{Department of Astrophysical Sciences, Princeton University, Princeton, NJ 08544, USA}
\affiliation{Center for Statistics \& Machine Learning, Princeton University, Princeton, NJ 08544, USA}
\email{peter.melchior@princeton.edu}
\author[0000-0002-8814-1670]{Sicong Lu}
\affiliation{Department of Physics and Astronomy, University of Pennsylvania, Philadelphia, PA 19104, USA}
\email{siconglu@berkeley.edu}
\author[0000-0001-6532-6755]{Quang H. Tran}
\affiliation{Department of Astronomy, Yale University,
New Haven, CT 06511, USA}
\email{quang.tran@yale.edu}

\begin{abstract}
The detection of Earth analogs with extreme-precision radial velocities (EPRVs) is limited by spectral variability from stellar activity, telluric absorption, and instrumental systematics. We apply \aestra, a generative spectrum modeling framework, to NEID Sun-as-a-star observations. \aestra empirically decomposes the spectra into stellar line-shape variability, micro-telluric absorption, and continuum variability without external atmospheric or stellar templates. After removing the learned telluric and continuum components,
we train a low-dimensional representation of the spectrum to infer activity-driven apparent RVs jointly with candidate Doppler signals. We evaluate the method with 500 single-planet injection--recovery tests spanning periods of 2.5--400 days and semi-amplitudes of $K=0.1$--$0.7$~m\,s$^{-1}$, calibrating the detection criterion to yield zero spurious detections. At this matched confidence level, \aestra recovers 238 injected planets, including 13 with $K<0.3$~m\,s$^{-1}$, whereas traditional CCF-based activity-indicator detrending recovers 9 planets and none below $K=0.5$~m\,s$^{-1}$.
\end{abstract}



\section{Introduction} 

The search for Earth-like planets in the habitable zones of Sun-like stars remains one of the central challenges of modern exoplanet science. Next-generation spectrographs have been engineered to achieve instrumental stability at the 0.1 \metersec\ level (ESPRESSO, \citealp{pepe2010espresso}; NEID, \citealp{schwab2016design}; EXPRES, \citealp{jurgenson2016expres}; HARPS3, \citealp{thompson2016harps3}), placing Earth-analog Doppler amplitudes formally within instrumental reach. In practice, however, the limiting factor is not photon noise or instrumental stability. Stellar surface variability, including oscillations, granulation, and magnetic activity, induces apparent RV shifts at the meter-per-second level. Additionally, terrestrial atmospheric absorption and instrumental variations lead to RV data in which planetary signals are entangled with multiple sources of structured spectral variability.


Telluric and instrumental effects are often treated in preprocessing steps before the stellar RV analysis. Telluric absorption is commonly handled either by masking contaminated wavelength regions or by forward-modeling the atmospheric transmission using line-by-line radiative-transfer calculations \citep[e.g.,][]{2005JQSRT..91..233C,2014A&A...564A..46B,2015A&A...576A..77S}, together with an assumed instrumental line-spread function (LSF). These approaches might not be effective enough for Earth-analog searches. Shallow micro-telluric lines with depths below the percent level can be difficult to identify or mask, and their correction depends critically on the accuracy of the molecular line data, wavelength solution, and adopted instrumental LSF, which are often imperfectly known \citep[e.g.,][]{2014A&A...568A..35C,2022A&A...666A.196A}. At sub-\metersec~precision, the instrumental LSF itself varies with wavelength and time, so a fixed LSF model per spectral order may not fully describe the instrumental imprint \citep{2024MNRAS.530.1252S}. In addition, imperfect blaze corrections and time-variable instrumental fluctuations can leave residual continuum structure that is not removed by standard normalization procedures.

To mitigate stellar variability, a common approach is to measure apparent RVs by cross-correlating the spectrum with a binary mask or spectral template, and then decorrelating the RV time series against traditional activity indicators \citep[e.g.,][]{2011A&A...528A...4B}. \textcolor{black}{These diagnostics include activity indicators derived from the cross-correlation function (CCF) as well as independent chromospheric activity tracers such as the Ca II H\&K emission-line index.}
The CCF can be interpreted as a high signal-to-noise average line profile, and indicators such as its width, depth, and bisector span provide useful summaries of stellar line-shape variability. These summaries are powerful but not necessarily complete: stellar surface activity can distort spectral line profiles in wavelength-dependent ways, and different spectral lines do not respond identically to the same activity state \citep{2020A&A...633A..76C}.

Another major line of development has focused on modeling correlated structure in the RV time series. Gaussian Process (GP) regression has become a widely adopted framework for capturing quasi-periodic activity signals through flexible covariance kernels \citep{2014MNRAS.443.2517H, 2015MNRAS.452.2269R,2023ARA&A..61..329A}. While effective in many cases, such approaches operate after the spectral data have been compressed into a one-dimensional time series, thereby discarding much of the spectral information that probes the physical origins of variability.

Recent efforts have moved upstream toward CCF-level and spectrum-level modeling to retain more of the information present in the spectra. 
At the CCF level, \citet{2021MNRAS.505.1699C} showed that the autocorrelation function of the CCF can be viewed as a shift-invariant representation of line-profile changes, and used its principal components to separate activity-driven RVs from true Doppler shifts. Line-by-line RV techniques go a step closer to the spectra by exploiting differential behavior among spectral lines to mitigate activity signals \citep{2020A&A...633A..76C,2022A&A...659A..68C}. 
At the spectrum level, data-driven approaches such as \textit{wobble} introduced empirical modeling of telluric absorption at the spectral level \citep{2019AJ....158..164B}. 
More recently, \citet{2024arXiv240817289G} jointly modeled stellar and telluric variability in stellar spectra as linear combinations of templates, retaining substantially more spectral information than CCF-based pipelines. In parallel, machine-learning approaches have been used to infer activity-driven RVs from CCF or reduced spectral ``shell'' representations, achieving sub-\metersec\ recoveries in HARPS-N solar data \citep{2022AJ....164...49D, 2024A&A...687A.281Z}.

In this work, we adopt a unified generative approach. We treat each observed spectrum as a composition of multiple variability sources and model these components jointly within a single framework. With this generative model, we learn a latent neural representation that captures correlated spectral changes across many spectral orders.
This representation plays a role analogous to traditional activity indicators: it provides a compact description of the stellar activity state that can be used to model activity-driven RVs. 
Unlike CCF-based indicators, however, the latent variables are learned directly from the spectra and are designed to capture differential line-profile variability across different wavelengths and spectral orders.
We then use this learned activity representation to separate the RV contribution associated with stellar line-shape variability from planetary Doppler shifts. 

\aestra was introduced by \citet{2024AJ....167...23L} and applied to simulated data.
This work presents the first application of 
\aestra to real data:
several years of Sun-as-a-star observations obtained with the NEID EPRV instrument mounted on the 3.5-meter WIYN telescope at Kitt Peak \citep{schwab2016design}. NEID’s dedicated solar feed delivers high-resolution, high-cadence spectra of the integrated solar disk on a daily basis \citep{2022AJ....163..184L}, with demonstrated long-term instrumental stability of $\approx$\,0.37~\metersec~\citep{2024arXiv240813318F}. The Sun provides a unique test for extreme-precision RV methodologies because it can be observed at high cadence with an exceptionally high signal-to-noise ratio. At the same time, solar spectra are affected by imperfect telluric correction, order-dependent instrumental systematics, and continuum variability, making this dataset a highly realistic challenge for spectrum-level modeling.

In this paper, we demonstrate that \aestra can empirically decompose real solar spectra into three dominant components: stellar activity–driven line-shape variability, micro-telluric absorption, and instrumental continuum fluctuations. By dividing out the learned telluric and continuum contributions while preserving stellar variability, we merge 42 high-quality NEID spectral orders into a single activity-dominated spectrum. We then train \aestra to learn a physically informative latent representation of stellar variability and use these latent features to infer the activity-driven contributions to the apparent radial velocities.


We also present a realistic validation method for Earth-analog searches by injecting synthetic planet signals into the NEID solar time series.
Because in a real RV survey the planet period is not known in advance, our recovery tests are period-blind: the analysis must select candidates without access to the injected parameters, count the planet as recovered only when the selected candidate matches the injected signal, and treat accepted signals at other periods as false positives.

The structure of the paper is as follows. In \autoref{sec:neid-data}, we describe the NEID solar observations and quality cuts used to construct the training sample. \autoref{sec:decompose} presents the spectral decomposition framework and validation tests. \autoref{sec:activity} introduces the activity-driven RV inference methodology, and \autoref{sec:injection} describes the injection–recovery analysis. We discuss limitations, failure modes, and extensions of \aestra in \autoref{sec:discussion}, and conclude in \autoref{sec:conclusion} with broader implications for generative spectral modeling in Doppler exoplanet surveys.



\section{NEID Solar Feed Data}
\label{sec:neid-data}

\begin{table*}[t]
\hspace*{-5em}
\begin{threeparttable}
\caption{Summary of data exclusions and quality cuts applied to the NEID Solar Feed spectra.}
\label{tab:data_cuts}
\small
\begin{tabular}{p{0.18\textwidth} p{0.45\textwidth} p{0.32\textwidth}}
\hline
\hline
Selection step & Criterion & Rationale \\
\hline
Fire shutdown$^{a}$ & Exclude 2022 Jun 13--2022 Nov 30 & No routine observations during shutdown\\
Hardware issues$^{a}$ & Exclude 2021 Oct 2--27; 2023 Feb 5--27; 2023 Mar 18--31 & Known instrumental anomalies \\
Solar eclipses$^{a}$ & Exclude 2023 Oct 14 and 2024 Apr 8 & Partial occultation of the solar disk \\
Airmass cut$^{a}$ & Reject exposures with airmass $> 2.5$ & Reduce high-airmass systematics \\
Cloud passages & Reject exposures within 15 min of irradiance drops $\geq 5\%$ & Remove transient cloud contamination \\
Low-count days & Reject days with $<10$ surviving spectra after cloud cuts & Exclude days with insufficient clean data \\
Low-irradiance days & Reject days with mean irradiance $<500$ W m$^{-2}$ & Exclude days dominated by poor weather \\
Intraday RV instability & Reject days with intraday RV RMS $>1.5$ m s$^{-1}$ & Unusually large within-day RV scatter \\
Large daily RV offset & Reject days with daily mean RV offset $>5$ m s$^{-1}$ & Anomalous day-to-day RV offsets \\

\hline
\end{tabular}
\begin{tablenotes}[flushleft]
\centering
\item[$^{a}$] Criterion adopted from \citet{2024arXiv240813318F}.
\end{tablenotes}
\end{threeparttable}
\end{table*}

\begin{figure*}[htbp]
    \centering
    \includegraphics[width=\textwidth,trim={2cm 1.4cm 2.7cm 2.5cm}, clip]{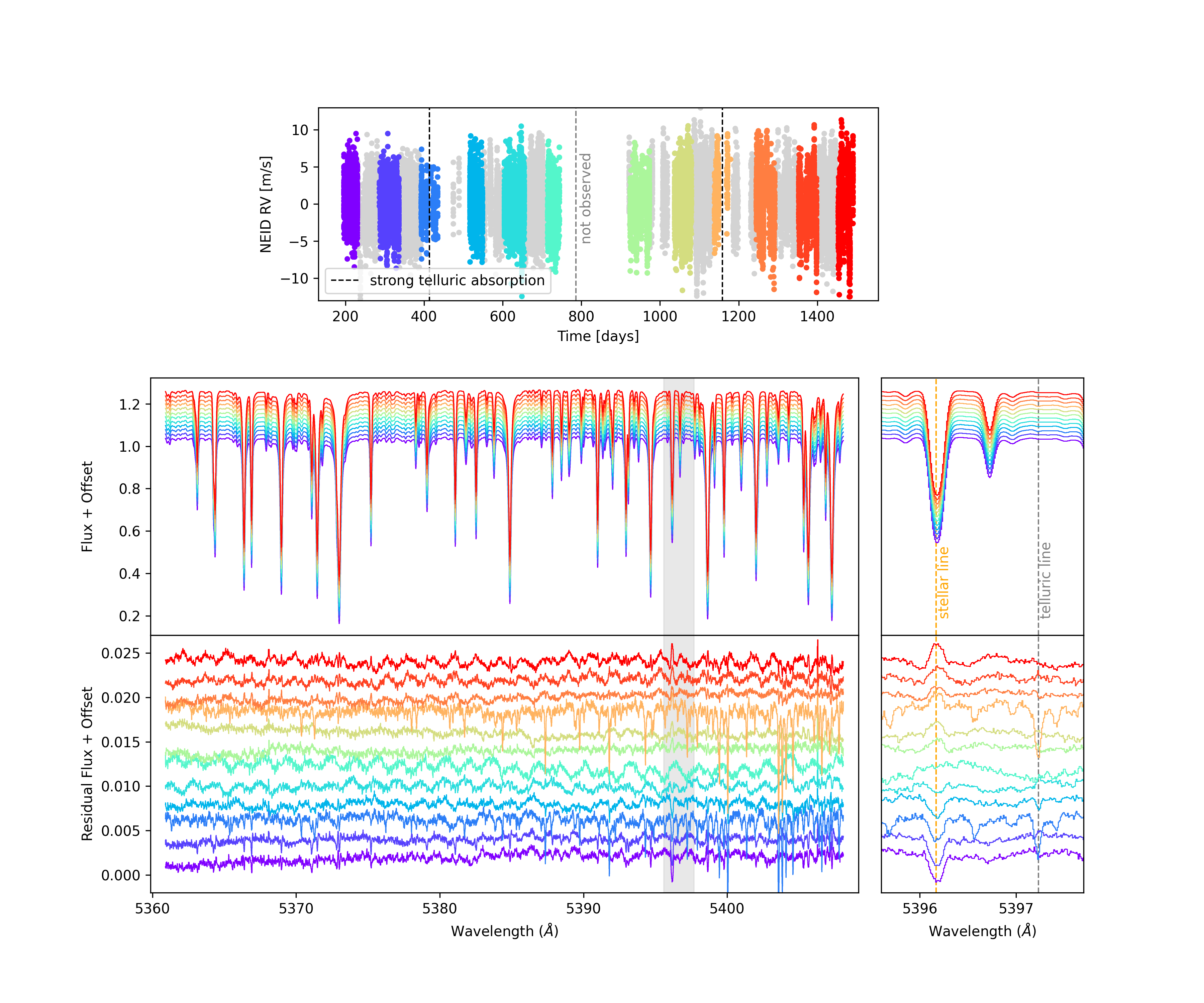}
    \caption{
        Raw spectral variability in NEID solar observations. \textit{Top:} CCF radial velocities from a representative spectral order versus time, with twelve selected observing segments highlighted (colored). Dashed vertical lines mark the seasonal phases of strong telluric absorption; the gray dashed line indicates the corresponding phase that was not observed during the fire-related shutdown. \textit{Middle:}  
        Weighted average spectra for these segments (vertically offset for clarity). \textit{Bottom:} Residual spectra relative to the global median spectrum, revealing low-amplitude but structured variability across wavelengths. \textit{Bottom Right:} Zoom of the shaded region showing a broad stellar absorption line (left dashed line) and a narrow telluric feature (right dashed line). 
        The two observed segments with the deepest telluric absorption are from similar times of the year. 
        The coexistence of broad stellar distortions, narrow atmospheric features, and continuum fluctuations illustrates the multi-scale nature of spectral variability that needs to be disentangled.}
    \label{fig:raw_spec}
\end{figure*}

\subsection{Dataset Overview and Instrumental RV Eras}


The NEID spectrograph is a fiber-fed optical echelle spectrograph \citep{schwab2016design,2019JATIS...5a5003R}. 
In its high-resolution mode, NEID provides optical wavelength coverage over approximately 380--930$\,{\rm nm}$ at a resolving power $R \gtrsim 100{,}000$. 
The spectrograph is housed in a vacuum chamber with sub-mK thermal stability and uses calibration sources including a laser frequency comb and a Fabry--P\'erot etalon to establish the wavelength solution and monitor instrumental drift.

The NEID Solar Feed directs disk-integrated sunlight from a small dedicated solar telescope into the stabilized spectrograph, enabling Sun-as-a-star observations \citep{2022AJ....163..184L}. 
During normal operations, the solar feed obtains 55-second exposures during daylight hours, corresponding to a cadence of roughly 1--2 minutes once detector readout is included.
These observations provide a densely sampled solar RV data set obtained with the same spectrograph used for nighttime stellar observations. 
Early commissioning observations demonstrated solar RV stability of 0.66~\metersec~under good sky conditions \citep{2022AJ....163..184L}, and subsequent analyses have reported long-term daily-averaged stability at the $\sim$0.37~\metersec~level \citep{2024arXiv240813318F}.

We obtained the publicly available NEID Solar Feed data from the Solar Radial Velocity Archive.\footnote{\url{https://neid.ipac.caltech.edu/search_solar.php}} We downloaded the Level-1 data products reduced with version 1.4.2 of the NEID data reduction pipeline (DRP). 
For each exposure, these wavelength-calibrated spectra contain 122 echelle orders, each sampled by 9216 pixels, along with corresponding flux uncertainties and wavelength solutions. 
We also downloaded accompanying Level-2 quantities, including pipeline RVs, activity indicators, water-vapor estimates, quality metadata, and pyrheliometer measurements, for quality control and validation.

A major interruption in observing occurred after the Kitt Peak wildfire in June 2022, which forced a suspension of NEID operations. 
When observations resumed in December 2022, the data belonged to a new instrumental RV era, with a documented RV zero-point offset and updated wavelength-calibration products, and possible mechanical or optical changes affecting the instrument response \citep{2024arXiv240813318F,NEIDDRP_RVEras}. For the purposes of spectral modeling, we therefore treat the pre-fire and post-fire observations as belonging to distinct instrumental states.
To enable joint modeling of spectra obtained before and after the fire, we introduce an empirical wavelength correction to align the pre-fire wavelength scale with that of the post-fire observations. The optimization of this correction is described in \autoref{subsec:wave}.

Regular solar observations were paused again on 2024 July 1, when the etalon calibration source became unavailable for monitoring instrumental drift. Although solar-feed observations resumed in 2025, we do not include data obtained after 2024 July 1 in this analysis, because the post-resumption baseline is still comparatively short and would not materially improve the present study.

\subsection{Quality Cuts}

We begin with the full set of 233,211 publicly available NEID solar spectra obtained between December 2020 and June 2024. From this parent sample, we apply a sequence of exclusions and quality cuts designed to remove observations affected by known instrumental problems, solar eclipses, poor weather, or anomalous RV behavior. The cuts are applied sequentially, and the percentages quoted below denote the fraction of the original parent sample removed at each stage, after all preceding cuts have already been applied. A summary of all cuts is provided in \autoref{tab:data_cuts}.

Following \citet{2024arXiv240813318F}, we first exclude intervals associated with known disruptions to normal operations, including the fire-related shutdown from 2022 June 13 through 2022 November 30, periods affected by documented hardware issues (2021 October 2--27, 2023 February 5--27, and 2023 March 18--31), and the dates of the partial solar eclipses on 2023 October 14 and 2024 April 8. Together, these date-specific cuts remove 2.8\% of the parent sample. We also reject all exposures taken through an airmass greater than 2.5 to reduce systematics associated with high-airmass observations, removing an additional 23.4\% of the parent sample. This cut mitigates both generic high-airmass effects relevant to stellar RV work, such as stronger telluric absorption, and solar-feed-specific effects from variable weighting of the resolved solar disk.

To remove exposures affected by transient cloud passages, we use the pyrheliometer data recorded by the NEID Solar Feed as a high-cadence monitor of direct solar irradiance. Rapid drops in irradiance indicate transient obscuration of the solar disk, which can induce artificial RV shifts through asymmetric illumination. We identify cloud passages as irradiance drops of at least 5\% and reject all exposures taken within 15 minutes of such events. 

We further reject days dominated by poor weather or insufficient clean data, excluding days with fewer than 10 surviving spectra after the cloud-based cuts or with mean irradiance below 500 W m$^{-2}$. For reference, clear days during the NEID solar-feed observing window, 16:30--22:30 UT typically have mean pyrheliometer irradiances of 800--1000 W m$^{-2}$. These weather-related cuts remove an additional 37.9\% of the parent sample.

Specifically, we reject any day for which the intraday root-mean-square scatter in the NEID CCF RVs exceeds 1.5 m s$^{-1}$ or for which the daily mean CCF RV differs from the global mean RV by more than 5 m s$^{-1}$. Among the days remaining after the preceding cuts, the median intraday RV RMS is 0.8 m s$^{-1}$, and the 1.5 m s$^{-1}$ threshold lies near the 95th percentile of the distribution. These cuts therefore remove only the high-scatter tail of days dominated by instrumental instability or other large-scale systematics, corresponding to an additional 4.8\% of the parent sample.

\subsection{Final Sample and Raw Spectral Variability}
After all exclusions and quality cuts, 72,449 high-quality spectra remain. Because full-model training on the entire set is computationally expensive, we randomly select a subset of approximately 30,000 spectra for subsequent analysis. This final sample spans a baseline of 1294 days and covers 521 clear observing days.

Before describing the preprocessing and modeling steps, it is useful to examine the raw spectral variability present in these data. The NEID solar spectra exhibit substantial time-dependent structure across multiple wavelength scales. \autoref{fig:raw_spec} illustrates this behavior for a representative wavelength region within a single echelle order. Composite spectra constructed from twelve representative observing segments appear nearly identical at first glance, but residuals relative to the global median reveal coherent, low-amplitude variability at the $\lesssim$ 1\% level across wavelengths. This variability spans multiple spatial scales, including broad continuum fluctuations, smooth distortions of stellar absorption lines, and sharp changes in narrow features associated with telluric absorption. In the zoomed region, a stellar line exhibits gradual changes in depth and shape over time, while a nearby telluric line varies more abruptly and independently. 
The strongest telluric absorption appears in two observed segments separated by approximately two years, both obtained at roughly the same time of year. 
The same seasonal phase in the intervening year fell during the fire-related shutdown and was not observed, so the pattern is consistent with annual modulation of atmospheric water-vapor absorption.
Thus, the astrophysical signal of interest is embedded in a mixture of broad continuum changes, narrow telluric features, and subtle stellar line-profile distortions, all sampled through an irregular observing window. In \autoref{subsec:preprocessing}, we describe the preprocessing and spectral decomposition procedure used to separate these contributions while preserving the stellar variability needed for downstream RV inference.

\begin{figure*}[htbp]
    \centering
    \includegraphics[width=\textwidth,trim={1cm 2cm 2cm 0cm}, clip]{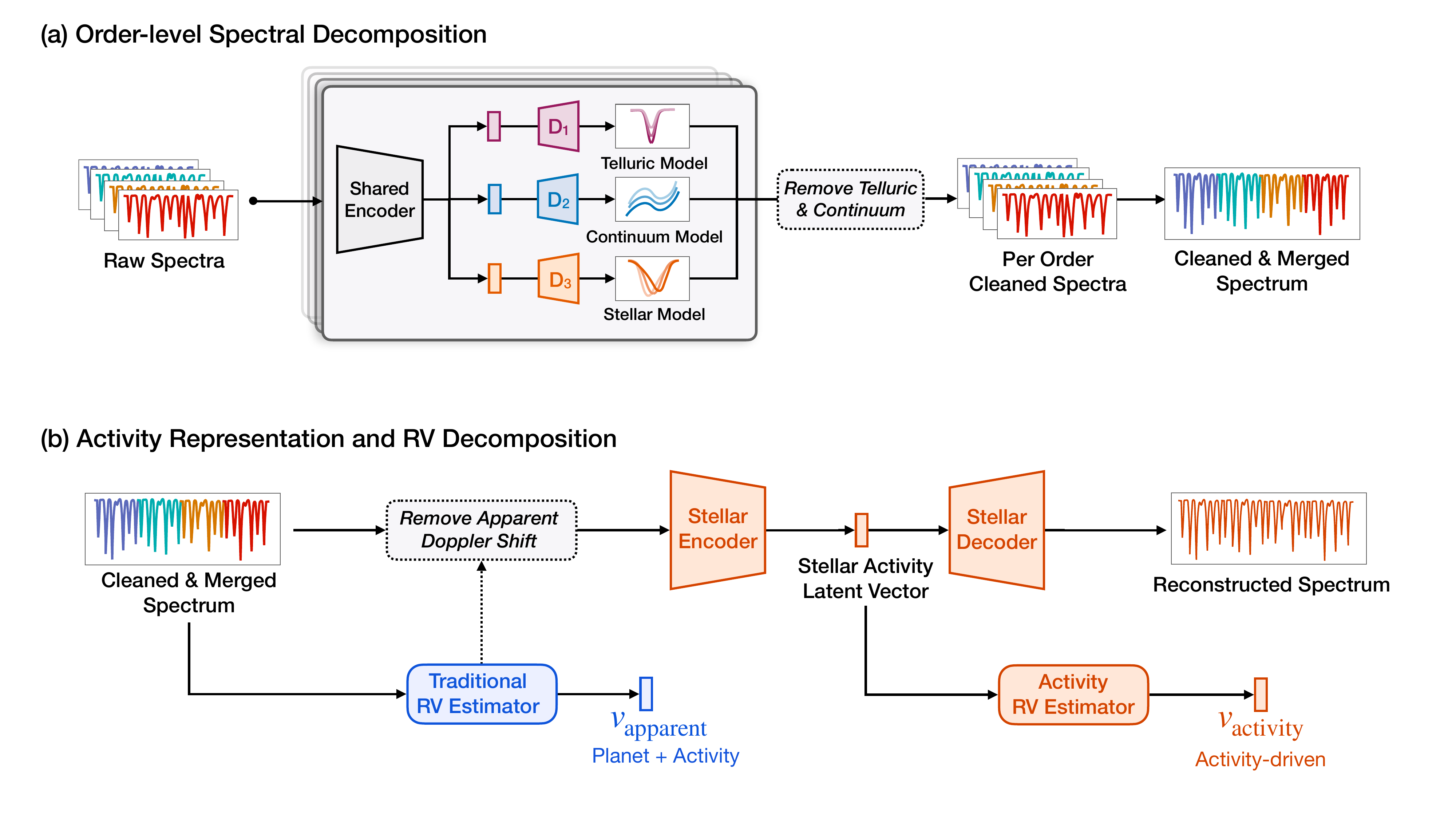}
    \caption{Overview of the \aestra\ workflow. (a) Each echelle order is decomposed into telluric absorption, continuum variability, and stellar line-shape variability using a shared encoder and component-specific decoder branches. The telluric and continuum components are removed, while stellar variability is preserved, yielding per-order cleaned spectra that are concatenated into a cleaned and merged spectrum. (b) The cleaned and merged spectrum is used to measure an apparent RV containing both activity and planetary contributions. After this apparent Doppler shift is removed, the spectrum is encoded into a low-dimensional stellar-activity latent vector. An activity RV estimator maps this latent representation to the activity-driven RV and is jointly optimized with candidate planetary signals, allowing coherent Doppler variability not explained by stellar line-shape variability to be modeled as planet candidates.
}
    \label{fig:architecture}
\end{figure*}

\section{Constructing Activity-Preserving Spectra}
\label{sec:decompose}
\subsection{Order Selection and Data Preprocessing}
\label{subsec:preprocessing}

We model the NEID solar spectra at the level of individual exposures and individual echelle orders, without temporal binning. The purpose of this preprocessing stage is to convert the extracted spectra into normalized, order-level inputs suitable for spectral decomposition on a common stellar-rest-frame wavelength grid.

Each NEID spectrum consists of 122 echelle orders, each sampled by 9216 pixels, covering a total wavelength range of 3621--10313\,\AA. This range extends slightly beyond the 380--930 nm range usually quoted for NEID, because the bluest and reddest extracted orders have low signal-to-noise ratios and are not generally useful for precision Doppler work. Because the signal-to-noise ratio is highest near the center of each order and declines toward the edges, where the wavelength solution is also less reliable, we restrict each order to its free spectral range using the NEID FSR mask\footnote{\texttt{neidMaster\_FSR\_Mask20210218\_v002.fits}, obtained by private communication.}. This mask removes the low-S/N order edges and overlapping regions, with the retained number of pixels depending strongly on order. After this masking, each order retains 3000--8000 spectral pixels, and the surviving wavelength intervals from neighboring orders can later be merged into a continuous one-dimensional spectrum.

We then select the wavelength range used for the spectral decomposition. After inspecting the raw spectral variability across all orders, we restricted the analysis to 4300--6230\,\AA, which corresponds to echelle orders 142–99 in NEID DRP notation. Bluer orders have lower signal-to-noise ratios and stronger continuum variability, while redder orders are increasingly dominated by telluric absorption, leaving less usable stellar Doppler information. This selection yields 44 candidate orders. After training the order-level decomposition model, we further rejected two orders spanning 5875--5990\,\AA, or echelle orders 103 and 104, where strong telluric absorption remains inadequately corrected (see details in \autoref{subsec:clean_merge}). The final analysis therefore uses 42 echelle orders.

For each retained order, we divide the science-fiber flux by the blaze function provided by the NEID Level 2 data product. The NEID DRP derives these blaze response vectors from continuum-lamp flat-field exposures, including a laser-driven light source (LDLS) for the blue orders and a quartz broadband lamp for the red orders. We then normalize the blaze-corrected spectrum by its median flux, so that the continuum level is approximately unity. The corresponding flux uncertainties are propagated through the same normalization. We also remove problematic pixels: pixels with negative flux values and isolated single-pixel artifacts are assigned zero flux and zero weight. The resulting normalized spectra and inverse-variance weights provide the order-level inputs to the decomposition model described below.

\subsection{Empirical Wavelength Alignment}
\label{subsec:wave}
After normalization and masking, spectra from each order must be placed on a common wavelength grid before they can be modeled jointly. We first apply the NEID barycentric correction to remove the dominant relative motion between the Earth and the target star. 
For the solar data analyzed here, RV contributions from known Solar System bodies are also removed.

For the Sun, the barycentric correction is especially well constrained, but in our analysis its role is only to place the spectra in a barycentric-corrected modeling frame, not to define an exact physical rest frame or remove any remaining astrophysical Doppler signals. The same step applies to exoplanet hosts: the barycentric correction removes the Earth's projected motion, while the stellar systemic velocity, activity-induced apparent shifts, and any planetary reflex motion remain in the data.

A further complication is that the wavelength solutions before and after the 2022 Kitt Peak wildfire are not fully consistent. To model the pre-fire and post-fire spectra within a single framework, we apply an empirical, low-order wavelength correction to the pre-fire spectra in each echelle order. Specifically, we use a smooth cubic-spline correction shared by all pre-fire spectra in that order to remove the dominant low-order mismatch relative to the post-fire wavelength solution. The corrected spectra and their inverse-variance weights are then interpolated onto a fixed stellar-rest-frame wavelength grid, producing the order-level inputs $(\mathbf{y}_{\mathrm{obs},i}, \mathbf{w}_{\mathrm{obs},i})$ used by the decomposition model below. The full mathematical form of the wavelength correction and interpolation is given in \autoref{appendix:wavelengths}.

\begin{figure*}[htbp]
    \centering
    \includegraphics[width=\textwidth,trim={0cm 0cm 0cm 0cm}, clip]{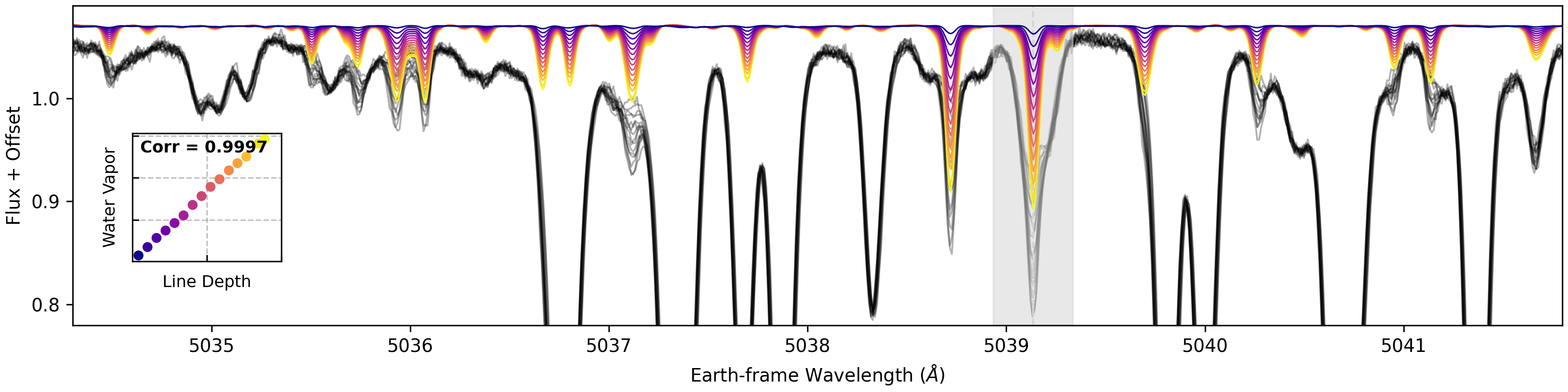}
    \caption{Example telluric components extracted by the spectral decomposition model. Gray curves show observed NEID solar spectra, while colored curves show the inferred telluric model, color-coded by the depth of the strongest extracted telluric line. The vertical gray shaded region marks this line, whose depth is used as an empirical proxy of telluric strength. The extracted features remain fixed in Earth-frame wavelength and vary mainly in depth, matching the expected behavior of terrestrial absorption. \textit{Inset:} The empirical telluric-strength proxy is tightly correlated with the NEID DRP best-fit precipitable water vapor value (PWV), with correlation coefficient $>0.999$. This agreement supports the interpretation that the extracted telluric component isolates atmospheric variability rather than stellar line-shape changes or continuum structure.
}
    \label{fig:tellurics}
\end{figure*}

\subsection{Per-order Decomposition into Telluric, Continuum, and Stellar Components}
\label{subsec:joint}

After preprocessing, we model each retained echelle order independently at the level of individual exposures. The goal of this stage is to separate the dominant non-astrophysical variability in the spectra---narrow telluric absorption and smooth continuum fluctuations---while preserving the stellar line-shape variability needed for downstream activity modeling and radial-velocity inference.

For each exposure $i$, we represent the observed spectrum in a given order as
\begin{equation}
\mathbf{y}_{{\rm model},i}
=
(1-\mathbf{y}_{{\rm t},i}) \odot (1+\mathbf{y}_{{\rm c},i}) \odot
\left(\bar{\mathbf{y}}_{\star} + \Delta \mathbf{y}_{\star,i}\right) + b_i,
\label{eq:order_model}
\end{equation}
where $\mathbf{y}_{{\rm t},i}$ describes telluric absorption, $\mathbf{y}_{{\rm c},i}$ describes smooth continuum variability, $\bar{\mathbf{y}}_{\star}$ is a trainable, time-independent stellar template for that order, $\Delta\mathbf{y}_{\star,i}$ captures time-variable stellar line-shape changes, and $b_i$ is a small scalar offset.  All bold quantities denote one-dimensional spectra defined on the common stellar-rest-frame wavelength grid $\bm{\lambda}_{\rm star}$, and $\odot$ denotes element-wise multiplication. The trainable template $\bar{\mathbf{y}}_{\star}$ is initialized from the average of the aligned, normalized spectra. Optimizing $\bar{\mathbf{y}}_{\star}$ jointly with the other components allows the model to absorb the static stellar spectrum into the shared template, so that the regularized stellar component represents only time-variable line-shape structure.

The model uses a shared encoder to compress each preprocessed spectrum into a low-dimensional latent representation, which is then decoded into component-specific branches for telluric, continuum, and stellar variability. The telluric component is constrained to describe narrow absorption-like structure, the continuum component captures smooth multiplicative variations, and the stellar component captures localized line-profile distortions.
The telluric component also includes a learned broadening kernel that empirically describes the line profiles of the narrow telluric absorption features in the extracted spectra. The decomposition model is optimized jointly with the empirical wavelength-correction spline described above, allowing pre-fire and post-fire spectra to be modeled within a single framework.
\autoref{fig:architecture} summarizes the overall workflow.

Because the decomposition is trained primarily through spectral reconstruction, the three branches would be partially degenerate without additional guidance. We therefore include weak structural constraints that separate the components by both morphology and wavelength-frame behavior: the telluric component describes narrow absorption features fixed in the Earth frame with exposure-dependent depths, the continuum component describes smooth low-frequency variations, and the stellar component describes broader, more flexible distortions localized around stellar absorption lines in the stellar frame. These constraints are not external atmospheric or stellar templates; they serve only as weak priors that guide the model toward physically interpretable decompositions. The full loss function, regularization terms, and hyperparameter choices are given in \autoref{appendix:loss}.

The learned telluric component recovers the expected behavior of terrestrial absorption. As shown in \autoref{fig:tellurics}, the extracted component consists of narrow features at the correct wavelengths whose depths vary from exposure to exposure. The model recovers these features directly from the NEID spectra, without using an external atmospheric transmission template. As an external comparison, we use the NEID DRP precipitable-water-vapor value (PWV; header keyword \texttt{WVAPOR}), which is obtained by fitting the pipeline's line-by-line radiative-transfer telluric model grid to selected deblazed spectral regions. The learned telluric component is tightly correlated with this DRP PWV diagnostic, supporting the interpretation that this component isolates genuine atmospheric variability.

\begin{figure*}[htbp]
    \centering
    \hspace*{-0.7cm}\includegraphics[width=1.02\textwidth,trim={0cm 0.1cm 0cm 0cm}, clip]{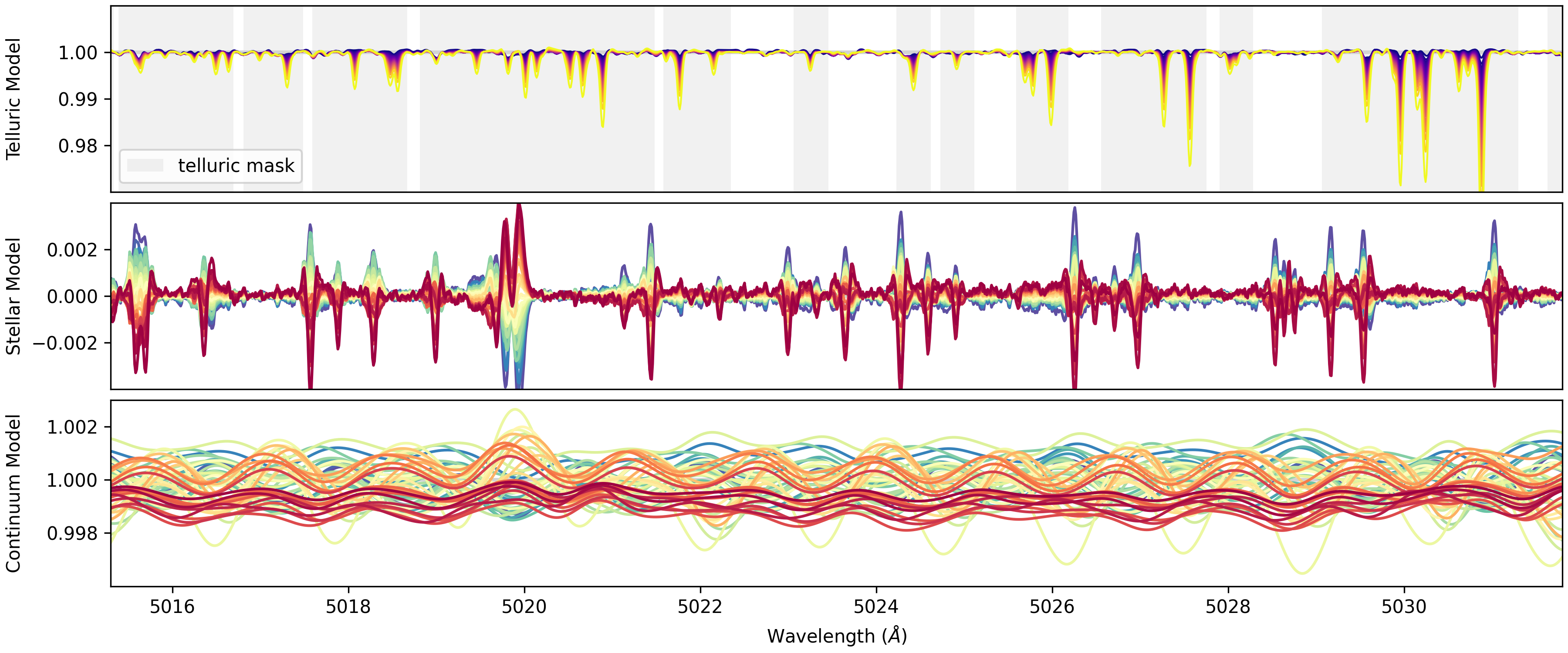}
    \caption{Examples of the spectral decomposition into three components capturing distinct sources of variability. \textit{Top:} The inferred telluric model consists of narrow absorption features; shaded regions indicate wavelengths identified by the telluric mask. Colors encode the empirical telluric strength, as defined in \autoref{fig:tellurics}. \textit{Middle:} The stellar model captures line-profile variations localized around spectral lines and is largely uncorrelated with the telluric mask, demonstrating separation from telluric absorption. Colors rank epochs by an empirical stellar activity proxy, defined as the depth of the line with the largest stellar variability in this order. \textit{Bottom:} The continuum model describes smooth, low-order multiplicative variations across wavelength, also color-coded by stellar activity level. Together, the three panels illustrate the separation of narrow telluric absorption, localized stellar line-shape variability, and broad continuum modulation within the same wavelength region.}
    \label{fig:order51_decomp_models}
\end{figure*}

A second diagnostic is whether the three branches converge to distinct spectral morphologies rather than arbitrarily partitioning the same residual structure. \autoref{fig:order51_decomp_models} shows that they do. The telluric component is sparse and line-like, with narrow absorption features whose depths vary across epochs. The continuum component is smooth and low-frequency, with no evident narrow-line structure. The stellar component is qualitatively different from both: it is localized around stellar absorption lines, where it captures asymmetric line-profile distortions. This clean morphological separation is a key empirical validation of the decomposition.

After training, we use only the learned telluric and continuum components to correct the spectra. The per-order stellar branch is not used as the stellar activity representation in the downstream analysis; its role is to absorb line-shape variability during decomposition, thereby preventing astrophysical structure from being mistakenly assigned to the telluric or continuum components. We therefore remove only the non-astrophysical telluric and continuum contributions from each order, as described in \autoref{subsec:clean_merge}, leaving the observed stellar line-shape variability in the corrected spectra. These activity-preserving per-order spectra are then merged and passed to the full-spectrum stellar activity model described in \autoref{sec:activity}.
 
\subsection{Cleaning and Merging the Activity-Preserving Spectra}
\label{subsec:clean_merge}
We now describe the construction of the activity-preserving spectra used in the downstream activity model. It involves three steps: applying the learned telluric correction only in wavelength regions with robustly determined atmospheric variability, dividing out the learned continuum component, and concatenating the retained orders onto a single merged wavelength grid.

We apply the telluric correction only in wavelength regions where the decomposition identifies a robust telluric signal. 
For each order, we first check the morphology of the learned broadening kernel associated with the telluric component. The kernel is initialized as a broad Gaussian. If training narrows the kernel and preserves a single dominant peak, we interpret the telluric component as having identified a localized absorption component and allow a telluric correction for that order. If the kernel instead broadens or develops multiple peaks, we regard that order as having no reliable telluric detection and apply no telluric correction.

For orders that pass this check, we use the depth of the strongest recovered telluric line as an empirical telluric-strength proxy, identify wavelength bins whose recovered telluric amplitudes correlate with this proxy at $r>0.9$, and mask $\pm 0.2$ \AA\ regions around those bins. This procedure restricts the telluric correction to wavelengths where the learned component behaves like real atmospheric absorption, reducing the risk of propagating weak or spurious telluric structure into the cleaned spectra. An example telluric mask is shown in the top panel of \autoref{fig:order51_decomp_models}.

For each exposure and each retained order, the model returns the telluric component $\mathbf{y}_{{\rm t},i}$. Before applying the correction, we set the telluric model to zero outside the masked regions. Denoting the resulting masked telluric component by ${\mathbf{y}'}_{{\rm t},i}$, the cleaned spectrum is defined as
\begin{equation}
\mathbf{y}_{{\rm clean},i}=
\frac{\mathbf{y}_{{\rm obs},i}- b_i}{(1-{\mathbf{y}}_{{\rm t},i}') \odot (1+\mathbf{y}_{{\rm c},i})},
\end{equation}
where $\mathbf{y}_{{\rm c},i}$ denotes the continuum component and $b_i$ denotes the scalar offset term. 

We also empirically inflate the spectral variances to account for pixel-level artifacts that are not fully captured by the pipeline uncertainties.
For each cleaned spectrum, we subtract the median cleaned spectrum and estimate a local residual variance within a six-pixel sliding window. 
Because real solar and telluric features are broadened by the instrumental line-spread function and extend over many pixels, this local residual variance is primarily sensitive to pixel-scale ``glitches'' rather than to astrophysical absorption lines.

We denote the resulting empirical variance-inflation term by $\bm \sigma_{{\rm glitch},i}^2$ and add it to the pipeline variance, $\bm \sigma_{{\rm obs},i}^2$. The updated variance is then
\begin{equation}
\bm \sigma_{{\rm clean},i}^2 = \bm \sigma_{{\rm obs},i}^2 + \bm \sigma_{{\rm glitch},i}^2,
\end{equation}
and the inverse-variance weights are updated accordingly. This empirical correction assigns lower weights to pixels with excess local residual structure, reflecting reduced confidence in those pixels.


After telluric and continuum correction, each retained echelle order provides an activity-preserving spectrum and corresponding inverse-variance weights on a common wavelength grid. We combine these orders by concatenating them in wavelength order. Small overlaps between adjacent orders are trimmed so that each wavelength pixel appears only once; in practice this removes only a few pixels at each order boundary.

We begin with 44 orders spanning DRP orders 142--99, selected from the 4300--6230 \AA\  region. Most of these orders contain negligible to moderate telluric features with depths of order $\lesssim 10\%$, which are adequately described by the low-dimensional telluric component used in the decomposition. We exclude two orders, 104 and 103, spanning approximately 5875--5990\,\AA, because this interval contains a dense set of strong telluric H$_2$O absorption lines around the Na~I~D region \citep[e.g.,][]{1993A&A...271..734L,2021MNRAS.502.4392L}. In the most contaminated epochs, individual telluric features in these orders reach depths of roughly 50\% of the total flux. In this regime, the present telluric model is not flexible enough to separate atmospheric absorption robustly from stellar structure across all epochs; modeling such heavily contaminated orders would require a more expressive telluric component, such as a higher-dimensional model with multiple eigenspectra rather than the single-eigenspectrum model used here.

After these order-level cuts and the quality cuts described in \autoref{sec:neid-data}, the final merged data set contains $N_{\rm spec}=29{,}352$ spectra from 42 echelle orders. For each exposure, the cleaned fluxes and inverse-variance weights are concatenated in the same wavelength order, producing merged spectra $\left(\mathbf{y}_{\mathrm{merge},i}, \mathbf{w}_{\mathrm{merge},i}\right)$, for $i=1,\dots,N_{\rm spec}$. The merged wavelength grid spans 4307.50--6231.70 \AA\  and contains 170{,}923 spectral bins. This product preserves stellar line-shape variability while removing the dominant telluric and continuum contributions, and provides the basis for constructing the activity-model input spectra.

\subsection{Removing Apparent Doppler Shifts}
\label{subsec:zero_apparent_rv_spec}

The cleaned and merged spectra defined in \autoref{subsec:clean_merge} have already been corrected for the dominant barycentric motion of the Earth and for the low-order mismatch between the pre-fire and post-fire wavelength solutions. However, they still contain exposure-to-exposure apparent Doppler shifts at the \metersec\ level. 
These shifts include any true Doppler shifts from planetary reflex motion and activity-induced apparent RV variations produced by changes in stellar line profiles. 
Separating these two contributions is the central goal of the downstream RV analysis.

Before modeling the activity-driven spectral distortions themselves, we remove the measured apparent RV, denoted by $v_{{\rm app},i}$, from each merged spectrum. This scalar velocity includes any bulk spectral translation present in the data, regardless of whether its physical origin is stellar activity, instrumental effects, or true Doppler motion.
Removing $v_{{\rm app},i}$ from the spectrum therefore removes the leading-order bulk wavelength translation from the spectra used to train the activity encoder.
The purpose is to make the encoder sensitive primarily to residual line-shape variability, rather than to coherent wavelength shifts. The measured apparent RV time series is retained separately to be decomposed into activity and planetary components in \autoref{subsec:RV_decomposition}. 

This construction also makes the large-scale injection--recovery experiments computationally feasible. It allows the spectral activity representation to be trained once and reused across all injected systems, while only the downstream RV-decomposition model is retrained for each injection. The reason is as follows.

Let $v_{{\rm app},i}$ denote the apparent RV measured from the original merged spectrum and $v^{\rm inj}_{{\rm planet},i}$ denote a synthetic planetary RV signal. If this signal were injected directly into the spectra, the measured apparent RV would become
\begin{equation}
v_{{\rm meas},i}^{\rm inj} \simeq v_{{\rm app},i} + v^{\rm inj}_{\rm planet,i}.
\end{equation}
Because this total apparent RV is then removed before the spectrum is passed to the stellar activity encoder, the injected Doppler shift is removed as well, provided that it is accurately captured by the apparent RV measurement. Thus, for a purely Doppler-like injection, the activity-model input spectra are approximately unchanged:
\begin{equation}
\mathbf{y}_{{\rm in},i}^{\rm inj} \simeq \mathbf{y}_{{\rm in},i}.
\end{equation}
We verified this approximation numerically by applying synthetic Doppler shifts to the spectra and remeasuring them with the template-fitting approach. The recovered shifts agreed with the injected shifts with an RMS discrepancy of $\ll 1\,{\rm cm\,s^{-1}}$, far below the velocity scale of the injected planets and the stellar RV variability. 
We therefore keep the zero-apparent-RV spectra, activity encoder--decoder pair, and latent variables fixed across all injection experiments, and introduce each injected planet only in the RV-decomposition stage.

To measure the apparent shift, we use a template-fitting procedure. We first compute the average of all merged spectra to define a high signal-to-noise template spectrum, ${\mathbf y}_{\rm temp}$, on the merged wavelength grid $\boldsymbol{\lambda}_{\rm merge}$. 

For each exposure $i$, we then determine the velocity shift $v_{{\rm app},i}$ by fitting a Doppler-shifted version of this template to the observed merged spectrum $\mathbf y_{{\rm merge},i}$. For each trial velocity $v$, the shifted template is evaluated as
\begin{equation}
\mathbf y_{\rm temp}^{(v)}={\rm Interp}\left[
\boldsymbol{\lambda}_{\rm merge};
\boldsymbol{\lambda}_{\rm merge}\left(1 + v/c\right),
{\mathbf y}_{\rm temp}
\right],
\end{equation}
where $\mathrm{Interp}\left[
\boldsymbol{x}_{\rm new};\boldsymbol{x}_{\rm old},\boldsymbol{y}_{\rm old}\right]$ denotes cubic-spline interpolation implemented with \texttt{scipy.interpolate}. The best-fit velocity is the value of $v$ that minimizes
\begin{equation}
\chi_i^2(v)=\sum_j w_{{\rm merge},ij}\left[
y_{{\rm merge},ij}-y_{{\rm temp},j}^{(v)}
\right]^2 ,
\end{equation}
where the sum is taken over the valid wavelength pixels. This interpolation allows the fitted shifts to be much smaller than a single detector pixel. We refer to the best-fit velocity $v_{{\rm app},i}$ as the apparent RV of exposure $i$. 

We then shift each merged spectrum by $-v_{{\rm app},i}$ and interpolate it onto a common wavelength grid. Denoting this grid by $\boldsymbol{\lambda}_{\rm in}$, we define
\begin{align}
\mathbf{y}_{{\rm in},i}&=
\mathrm{Interp}\!\left(
\boldsymbol{\lambda}_{\rm in};
\, \boldsymbol{\lambda}_{\rm merge}\!\left(1-v_{{\rm app},i}/c\right),\, \mathbf y_{{\rm merge},i}\right), \\
\mathbf{w}_{{\rm in},i}&=
\mathrm{Interp}\!\left(
\boldsymbol{\lambda}_{\rm in};
\, \boldsymbol{\lambda}_{\rm merge}\!\left(1-v_{{\rm app},i}/c\right), \, \mathbf w_{{\rm merge},i}\right),
\end{align}
where $c$ is the speed of light and $\mathrm{Interp}$ again denotes cubic-spline interpolation. 
After interpolation, invalid pixels with negative fluxes or negative weights are masked by setting both the flux and weight to zero.
We refer to
$(\mathbf{y}_{{\rm in},i}, \mathbf{w}_{{\rm in},i})$ as the
activity-model input spectra.

\section{Stellar Activity Modeling and RV Inference}
\label{sec:activity}
The RV-inference strategy used here differs from the original \aestra framework described in \cite{2024AJ....167...23L}. In the original framework, the neural network was trained to estimate Doppler shifts directly from spectra. \textcolor{black}{In this work, apparent velocities have already been measured from the cleaned and merged spectra using the template-fitting procedure described in \autoref{subsec:zero_apparent_rv_spec}}, primarily for computational scalability. The tradeoff is that the present implementation does not yet test a fully integrated neural RV estimator on each injection experiment.

\subsection{Learning a Low-Dimensional Representation of Stellar Variability}
\label{subsec:activity_latent}

The activity-model input spectra constructed in \autoref{subsec:zero_apparent_rv_spec} are used to learn a compact representation of full-spectrum stellar line-shape variability. The stellar activity modeling and downstream RV analysis are summarized in \autoref{fig:architecture}. For each exposure $i$, the input consists of the spectrum $\mathbf{y}_{{\rm in},i}$ and inverse-variance weights $\mathbf{w}_{{\rm in},i}$ on the common empirical wavelength grid $\bm{\lambda}_{\rm in}$. These spectra have been cleaned of the dominant telluric and continuum contributions and shifted into a common input frame, but they retain the residual line-profile distortions associated with stellar variability.

We keep the spectra at the individual-exposure level rather than applying daily binning or averaging. In traditional RV analyses, multiple exposures within a night are often averaged to reduce the contribution of short-timescale stellar variability, especially acoustic oscillations and granulation. Here, the objective is different: the activity model is trained on the spectral features of each exposure. Daily binning would mix spectra obtained at different activity states and suppress some of the line-shape variability that we want the model to learn. Instead, these short-timescale spectral variations are retained in the training set and must be described by the learned stellar latent variables.


We model the stellar activity perturbations in input spectra using \aestra \citep{2024AJ....167...23L}, which implements an attentive autoencoder architecture introduced by \cite{melchior2023autoencoding}.
The encoder maps each input spectrum to an $S$-dimensional latent vector,
\begin{equation}
    \mathbf{z}_i = \mathrm{Encoder}(\mathbf{y}_{{\rm in},i}),
\end{equation}
which provides a compact empirical description of the stellar activity state of exposure $i$. The decoder maps this latent vector back to a full spectrum,
\begin{equation}
    \hat{\mathbf{y}}_{{\rm in},i} = \mathrm{Decoder}(\mathbf{z}_i),
\end{equation}
and the model is trained to reconstruct the input spectra as closely as possible using the inverse-variance weights. We intentionally keep the decoder simple: it consists of a single wide projection from the latent space to the spectral dimension, followed by an activation function. \textcolor{black}{This differs from the implementation introduced by \citet{2024AJ....167...23L}, which used a multi-layer decoder. The simplified architecture is motivated primarily by computational considerations: the merged spectra contain 170,923 wavelength bins, making a deeper fully connected decoder memory-intensive and unnecessarily flexible for the present application. The simpler decoder also reduces the risk of fitting pixel-scale noise and improves interpretability, because the learned latent variables map more directly onto global spectral perturbation patterns. Despite this reduced flexibility, we find that the reconstruction quality remains sufficient for the downstream activity-RV analysis.}

We found that the behavior of the model is mostly insensitive to the choice of the number of dimensions of the latent space over the range $S=5$--$20$, with $S=20$ giving slightly better reconstruction and performance. In the results below, we use $S=20$. 

This architecture imposes a strong bottleneck. Each input spectrum contains 170,923 wavelength bins, while the activity state of each exposure is described by only $S=20$ latent parameters. The model therefore cannot freely encode arbitrary pixel-level residuals or exposure-specific noise. Instead, it is forced to capture spectral variability that is repeatable and predictive across wavelengths. We interpret the resulting latent vector,
$\mathbf{z}_i$, as a low-dimensional representation of
the stellar activity state, which is used below to infer the activity-driven
RV contribution.

\begin{figure}[t!]
    \centering
    \hspace*{-0.7cm}\includegraphics[width=0.45\textwidth,trim={0cm 0cm 0cm 0.0cm}, clip]{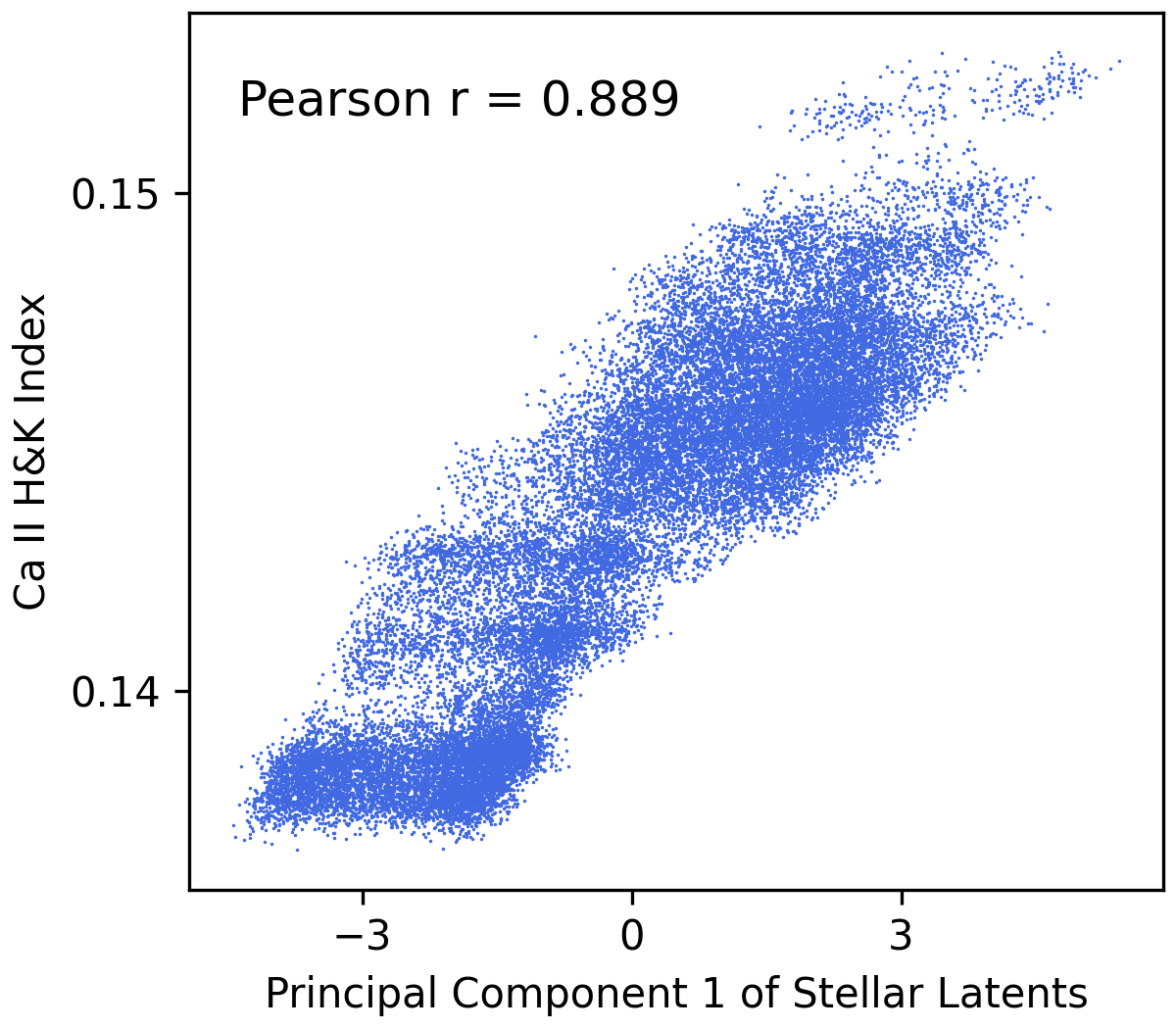}
    \caption{Correlation between the leading stellar latent component and an independent chromospheric activity tracer. The first principal component of the learned latent representation is strongly correlated with the NEID Ca II H\&K index ($r=0.889$), despite the Ca II H\&K region being excluded from training. This out-of-sample agreement indicates that the latent space captures physically meaningful stellar activity variability.}
    \label{fig:pc1_hk}
\end{figure}

To assess whether the learned latent representation captures physically meaningful stellar variability, we compare its leading modes to independent activity diagnostics. We focus first on the Ca II H\&K S-index reported in the \texttt{ACTIVITY} extension of the NEID Level 2 data products, following the S-index definition of \citet{1991ApJS...76..383D}. The Ca II K and H resonance lines are located at approximately 3933.7 and 3968.5$\,$\AA, respectively, and their line cores are sensitive to emission from magnetically heated chromospheric regions.

\autoref{fig:pc1_hk} shows that the first principal component of the stellar latent vectors is strongly and monotonically correlated with the Ca II H\&K index, with a Pearson correlation coefficient of $r=0.889$. This comparison is particularly informative because the Ca II H\&K region lies blueward of the wavelength range used to train the activity model. The correlation therefore provides out-of-sample evidence that the latent space is not merely encoding arbitrary spectral variation, but is recovering a dominant axis of real solar activity. 

At the same time, the relation is not perfectly one-dimensional. Principal Component 1 of stellar latent vectors explains only $\sim$20\% of the total variance in the stellar latent space, and the finite width of the correlation indicates that spectra with similar Ca II H\&K levels can still differ in other aspects of their line-shape variability. This is consistent with the expectation that stellar activity is not fully described by any single scalar indicator. Rather, the latent space appears to capture a richer, higher-dimensional description of the photospheric variability imprinted across the spectrum. Additional comparisons between the learned latent representation, traditional CCF-based activity indicators, and the pre-fire/post-fire subsets are presented in \autoref{appendix:latent}.

In the next subsection, these latent variables are mapped to an activity-driven radial-velocity contribution and used to separate stellar variability from Keplerian signals.

\subsection{Decomposition of Radial Velocity into Activity and Planetary Components}
\label{subsec:RV_decomposition}
Having learned a low-dimensional representation of stellar variability, we next use the stellar latent vector $\mathbf{z}_i$ to infer the activity-driven radial-velocity perturbation for each observation. The underlying assumption is that stellar activity produces RV shifts through spectral line-shape distortions, and that these distortions are encoded in the latent representation learned from the spectra. 

We introduce an \textit{Activity RV Estimator}, a small feedforward neural network that maps the stellar latent vector to a scalar activity-induced contribution to the apparent RV,
\begin{equation}
v_{{\rm act},i} = f_\theta(\mathbf{z}_i).
\end{equation}
In our implementation, $f_\theta$ is a multilayer perceptron with two hidden layers of sizes 128 and 8. This low-capacity architecture allows a flexible mapping from spectral state to RV while discouraging the activity model from fitting arbitrary time-dependent structure.

The remaining coherent RV variability is modeled with a separate time-dependent module, the \textit{Planetary RV Module}. We represent the planetary contribution as a sum of sinusoidal signals,
\begin{equation}
\label{eq:planets}
v_{{\rm planets},i}=\sum_{k=1}^{N_{\rm p}}K_k
\sin\left( \frac{2\pi t_i}{P_k} + \phi_k \right),
\end{equation}
where each component is defined by a semi-amplitude $K_k$, period $P_k$, and phase $\phi_k$. This model should be interpreted as a circular-orbit approximation to the Keplerian signal.
For mildly eccentric orbits, the Keplerian RV curve can be viewed approximately as a sinusoid at the orbital period plus eccentricity-dependent harmonics. We therefore expect a circular model to retain sensitivity to the dominant orbital period for low-eccentricity planets. Quantifying this effect requires dedicated eccentric injection--recovery tests, which we leave for future work.

The decomposition exploits the fact that stellar activity and planetary motion imprint differently on the spectra. Planetary signals primarily produce wavelength-independent Doppler shifts with deterministic temporal
behavior, whereas stellar activity generates wavelength-dependent line-profile variability that is not necessarily sinusoidal in time. By shifting each spectrum by $v_{{\rm app},i}$ before training the stellar encoder, we suppress bulk Doppler information and encourage the latent representation to encode residual spectral morphology rather than pure wavelength shifts. The activity RV estimator therefore predicts the component of the apparent RV associated with stellar line-shape variability, while coherent RV signals not explained by the learned activity representation are assigned to the planetary component.

This separation is not assumed to be exact. Activity signals can remain partially degenerate with sinusoidal signals, particularly when their characteristic timescales overlap or when the observing window introduces aliases. In practice, the optimization therefore depends on careful initialization and empirical calibration. In the next section, we describe the training strategy and injection--recovery framework used to stabilize the decomposition and evaluate detection reliability.

\section{Injection and Recovery Tests}
\label{sec:injection}

\subsection{Planet Injection Design}
To quantify the detection performance of our method, we conduct a suite of injection--recovery experiments on the NEID solar time series. These tests are designed to evaluate the sensitivity and reliability of the downstream activity-RV inference and planet-detection procedure, based on the spectral representation learned from the original data. 

As established in \autoref{subsec:zero_apparent_rv_spec}, a pure Doppler injection followed by apparent-RV measurement and removal leaves the activity-model input spectra unchanged to numerical precision. We therefore perform the injection experiments at the scalar-RV level: the zero-apparent-RV spectra, trained activity encoder--decoder, and latent vectors are held fixed, while each injected planet is added only to the apparent-RV time series used in the RV decomposition.

For each experiment, we construct an injected apparent-RV time series by adding a single synthetic planetary signal to the original apparent RVs:
\begin{equation}
v_{{\rm meas},i}^{\rm inj} = v_{{\rm app},i} + K_{\rm inj} \sin\left(2\pi t_i/P_{\rm inj} + \phi_{\rm inj}\right),
\end{equation}
where \(v_{{\rm app},i}\) denotes the apparent RV time series measured from the original, non-injected spectra after the preprocessing described above. We then retrain only the \textit{Activity RV Estimator} and \textit{Planetary RV Module} for each injected system, using the fixed latent activity vectors together with the injected apparent RVs. The subsequent detection criteria are applied uniformly to each experiment, without case-by-case tuning.

The injected signals are circular-orbit sinusoids spanning a broad range of orbital parameters: periods are drawn log-uniformly between \(2.5\) and \(400\, {\rm d}\), semi-amplitudes uniformly between \(K = 0.1\) and \(0.7\) \metersec, and orbital phases \(\phi_{\rm inj}\) are drawn uniformly from $0$ to $2\pi$. In total, the injection suite consists of 500 experiments.

\subsection{Traditional RV-Level Candidate Search}
\label{subsec:traditional_candidate_search}

Before applying the \aestra velocity decomposition, we perform a traditional RV-level period search on each injected velocity time series. This step serves two purposes. First, it provides an overcomplete set of candidate planetary periods used to initialize the planetary RV module in the \aestra training. Second, the same RV-level analysis, with an independently calibrated false-alarm probability (FAP) threshold, defines the traditional baseline against which we compare \aestra in \autoref{subsec:injection_performabce}.

The traditional search begins from the apparent RV time series measured from the same cleaned and merged spectra used by \aestra.
As a standard RV-level activity mitigation baseline, we model the apparent RVs as a linear function of CCF diagnostics commonly used to trace stellar line-profile variability \citep[e.g.,][]{2011A&A...528A...4B,2014ApJ...796..132D}. Specifically, we regress the apparent RVs against the CCF depth, FWHM, BIS SPAN, and a scalar offset term. The fitted activity contribution is subtracted to produce a detrended RV time series.

Candidate periods are constructed through a comprehensive period search on the detrended RV time series. We search multiple subsets of the data, including segments separated by the NEID wildfire shutdown and shorter contiguous windows to improve sensitivity to weak short-period signals that may be diluted or intermittently masked in the full time series. Full details of this period search procedure are provided in \autoref{appendix:period_search}.

After collecting all candidate periods from these searches, we fit the full detrended RV time series with a sinusoidal RV model initialized at each candidate period,
optimizing the period, semi-amplitude, and phase. We then sort the resulting candidate signals and merge near-duplicates, treating neighboring solutions as distinct only if their fractional period difference exceeds 10\%. This procedure usually results in 20--30 planet candidates per experiment.

This search is intended to provide a comprehensive list of plausible candidates for the subsequent joint activity--planet decomposition.
To test whether this proposal stage limits the present injection--recovery results, we measured how often the injected period appears among the top 20 initial candidates.
For injected signals with $K \sim 0.4$ \metersec, the true period is present in the candidate list in approximately 95\% of trials; for $K \sim 0.3$ \metersec\  and $K \sim 0.2$ \metersec, the corresponding fractions are approximately 80\% and 60\%, respectively. These fractions are substantially higher than the final validated detection completeness at the same amplitudes, indicating that most missed detections are rejected by the activity--planet decomposition and detection-quality criterion rather than being absent from the initial period proposal set.

\subsection{\aestra Initialization and Training}
\label{subsec:aestra_initialization}

The \aestra velocity decomposition is initialized from the candidate periods identified by the traditional RV-level search described in \autoref{subsec:traditional_candidate_search}. These candidates provide an overcomplete set of possible planetary periods, while the final planetary amplitudes and activity contribution are optimized within the \aestra model. In this sense, \aestra inherits the leading-order information captured by these indicators through the initial candidate set, while subsequently refining and re-evaluating these signals.

The activity and planetary components are fit through the velocity decomposition loss:
\begin{equation}
L_{{\rm vel},i}=
\left[v_{{\rm meas},i}^{\rm inj}- v_{{\rm act},i} - v_{{\rm planets},i} - v_{{\rm trend},i} \right]^2 ,
\end{equation}
where $v^{\rm inj}_{\rm meas,i}$ is the injected apparent RV of observation $i$, $v_{{\rm act},i}=f_\theta(\mathbf{z}_i)$ is the activity-driven RV inferred from the learned stellar latent variables, and $v_{{\rm planets},i}$ is the sum of initialized sinusoidal candidate signals. The term $v_{{\rm trend},i}$ denotes an optional preliminary correction based on classical activity indicators; when no such correction is applied, we set $v_{{\rm trend},i}=0\,$\metersec.

To reduce degeneracies in the joint optimization, we train separate decompositions for four period ranges: $P<10\,{\rm d}$, $10\,{\rm d}<P<100\,{\rm d}$, $100\,{\rm d}<P<250\,{\rm d}$, and $P>250\,{\rm d}$. For each period range, the planetary RV term ($v_{{\rm planets},i}$ in \autoref{eq:planets}) includes the full set of candidate signals identified in that range by the initial RV-level search. The periods and phases of these candidate signals are fixed to their initial best-fit values, while their semi-amplitudes are optimized jointly with the parameters of the activity RV estimator $f_\theta$. 

The purpose of this block-wise strategy is to allow candidate signals within the same period range to compete during training. Rather than fitting each candidate independently, which can assign significance to multiple aliases or redundant periodicities, the joint fit can suppress unnecessary candidates by driving their amplitudes toward zero while retaining the subset of candidates that best explains the data. This block-wise strategy also provides a computationally efficient way to explore a large candidate space without requiring repeated nonlinear optimization over period and phase.

The term $v_{{\rm trend},i}$ is not required by the \aestra formalism, but is used here as an optional preliminary correction step for the velocity decomposition.
The treatment of $v_{{\rm trend},i}$ depends on the period range. 
For the short- and intermediate-period runs, we estimate $v_{{\rm trend},i}$ with a multilinear regression against standard activity indicators and subtract it within the decomposition loss. This removes a leading-order RV-level activity component and reduces the dynamic range that must be modeled by the activity RV Estimator before fitting lower-amplitude coherent signals.

For the long- and extra-long-period runs, we instead set $v_{{\rm trend},i}=0~{\rm m~s^{-1}}$ and train directly on the apparent RVs, because linear detrending against activity indicators can more easily absorb or attenuate a long-period Doppler signal.

\subsection{Detection Rules and Calibration}
\subsubsection{\aestra Detection Statistic}
\label{subsubsec:aestra_detection}
The velocity decomposition returns a set of candidate periodic signals with optimized semi-amplitudes. We adopt a set of detection criteria to reject spurious or weakly constrained solutions. These criteria are defined once and applied uniformly across all recovery tests.

We first discard candidates with recovered semi-amplitude
$K_{\rm rec}<0.2$ \metersec. 
This amplitude cut represents an empirical sensitivity floor for the present analysis with the NEID solar data: candidates below this level did not correspond to robust recoveries and mainly introduced low-amplitude fitted noise structure into the planet-subtracted RVs used for the residual-periodogram test.
For the remaining candidates, we construct a planet-subtracted RV time series,
\begin{equation}
v_{{\rm resid},i} =
v_{{\rm meas},i}^{\rm inj} - v_{{\rm trend},i} - v_{{\rm planets},i},
\end{equation}
where $v_{{\rm planets},i}$ is the sum of all remaining planetary components in the corresponding training run. 
We intentionally do not subtract the activity term $v_{{\rm act},i}$, because the purpose of this diagnostic is to identify significant, activity-driven periodic power that remains after subtracting the candidate planetary components.

We then measure the residual Lomb--Scargle power at each candidate period. For candidates with $P<120\,{\rm d}$, the residual periodogram is computed separately for the pre- and post-shutdown data segments, and we use the larger of the two powers. For candidates with $P>120\,{\rm d}$, the residual periodogram is computed using the full time series.

We define a detection quality score that combines the recovered signal amplitude with the residual periodogram power,
\begin{equation}
Q_{\rm detect} = K_{\rm rec} - a \mathcal{P}_{\rm resid},
\end{equation}
where $K_{\rm rec} $ is the recovered semi-amplitude in ${\rm m\,s^{-1}}$, $\mathcal{P}_{\rm resid}$ denotes the dimensionless residual periodogram power at the candidate period, and we adopt a fixed normalization constant $a=3.5$. This numerical factor is chosen to place the amplitude term and residual-power term on similar scales. This statistic favors candidates with large recovered amplitudes while penalizing solutions for which significant power remains in the planet-subtracted time series.
This residual-power penalty is intended to reject candidates whose periodicity is not well described by a coherent sinusoidal signal, especially quasi-periodic activity signals whose amplitudes or phases evolve over time.

\begin{figure}[htbp]
    \centering
    \hspace*{-0.7cm}\includegraphics[width=0.45\textwidth,trim={0cm 0.0cm 0cm 0cm}, clip]{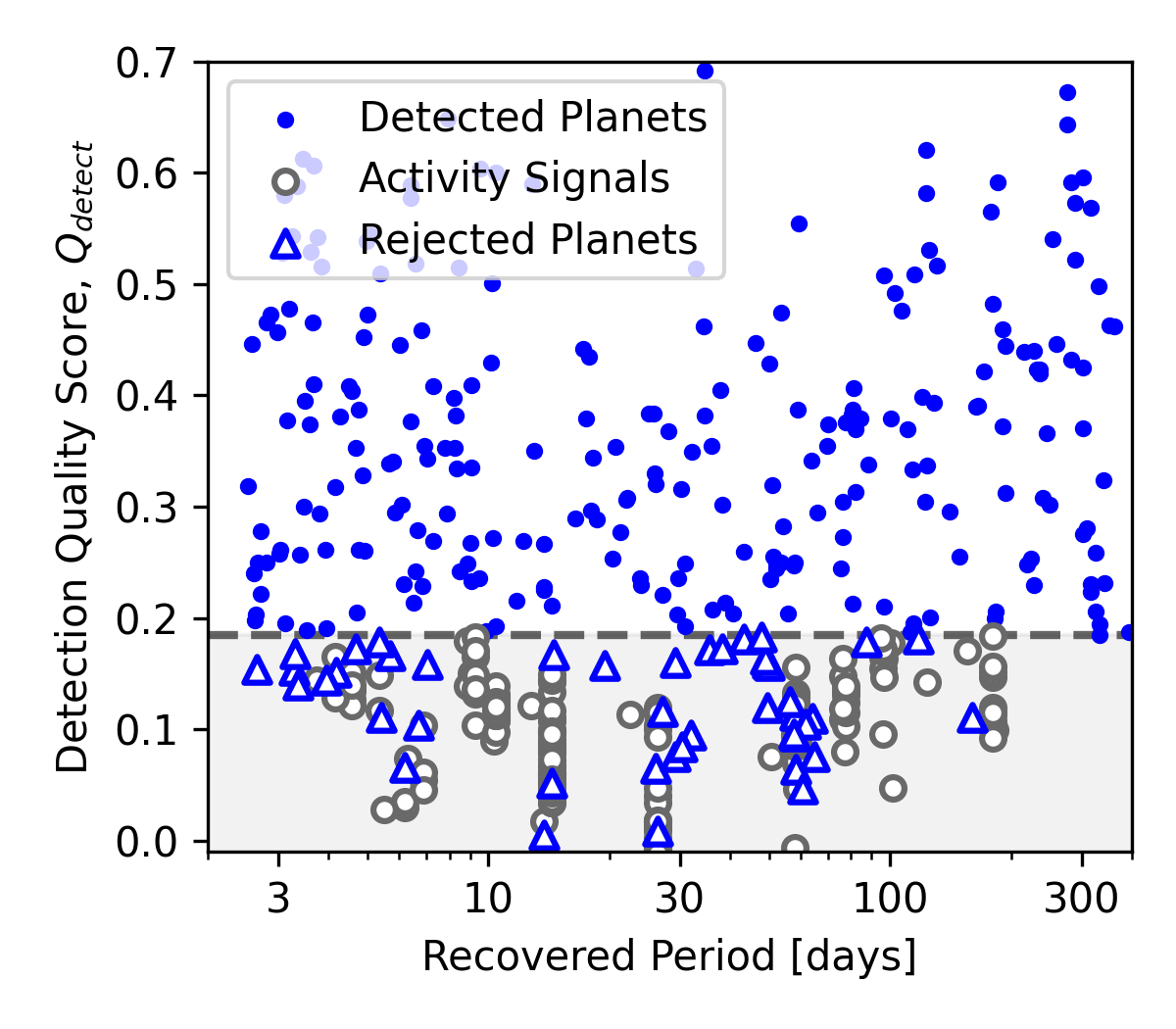}
    \caption{Detection quality score for candidate signals in the injection–recovery experiments. Blue circles mark correct recoveries. Blue triangles mark cases whose top-ranked candidates match the injected planets but do not pass the detection threshold. Gray circles mark top-ranked candidates that do not pass the detection threshold and do not match the injected signal. The dashed horizontal line indicates the fixed decision threshold, $Q_{\rm detect}=0.185$, applied uniformly to all trials.
    }
    \label{fig:aestra_decision}
\end{figure}

Candidates are accepted when $Q_{\rm detect}>0.185$, as shown in \autoref{fig:aestra_decision}. This threshold was calibrated on the injection--recovery suite by increasing the cutoff until no false positives were recovered among the 500 trials. After this calibration, the same threshold was applied to all experiments. The resulting rule defines the operating point used for the completeness and false-positive comparisons below.

As shown in \autoref{fig:aestra_decision}, the adopted criterion separates most recovered planetary signals from activity-induced or spurious periodicities. Successfully recovered injected planets lie predominantly above the boundary, while activity-induced signals cluster below it. Although this calibration uses injected signals with known ground truth, it should be possible to use the same strategy on real stars in exoplanet surveys. Synthetic planetary signals can be injected into the observed time series and processed through the full pipeline, allowing the detection boundary to be calibrated against controlled recoveries for that specific dataset.

\begin{figure}[htbp]
    \centering
    \hspace*{-0.7cm}\includegraphics[width=0.45\textwidth,trim={0cm 0.5cm 0cm 0cm}, clip]{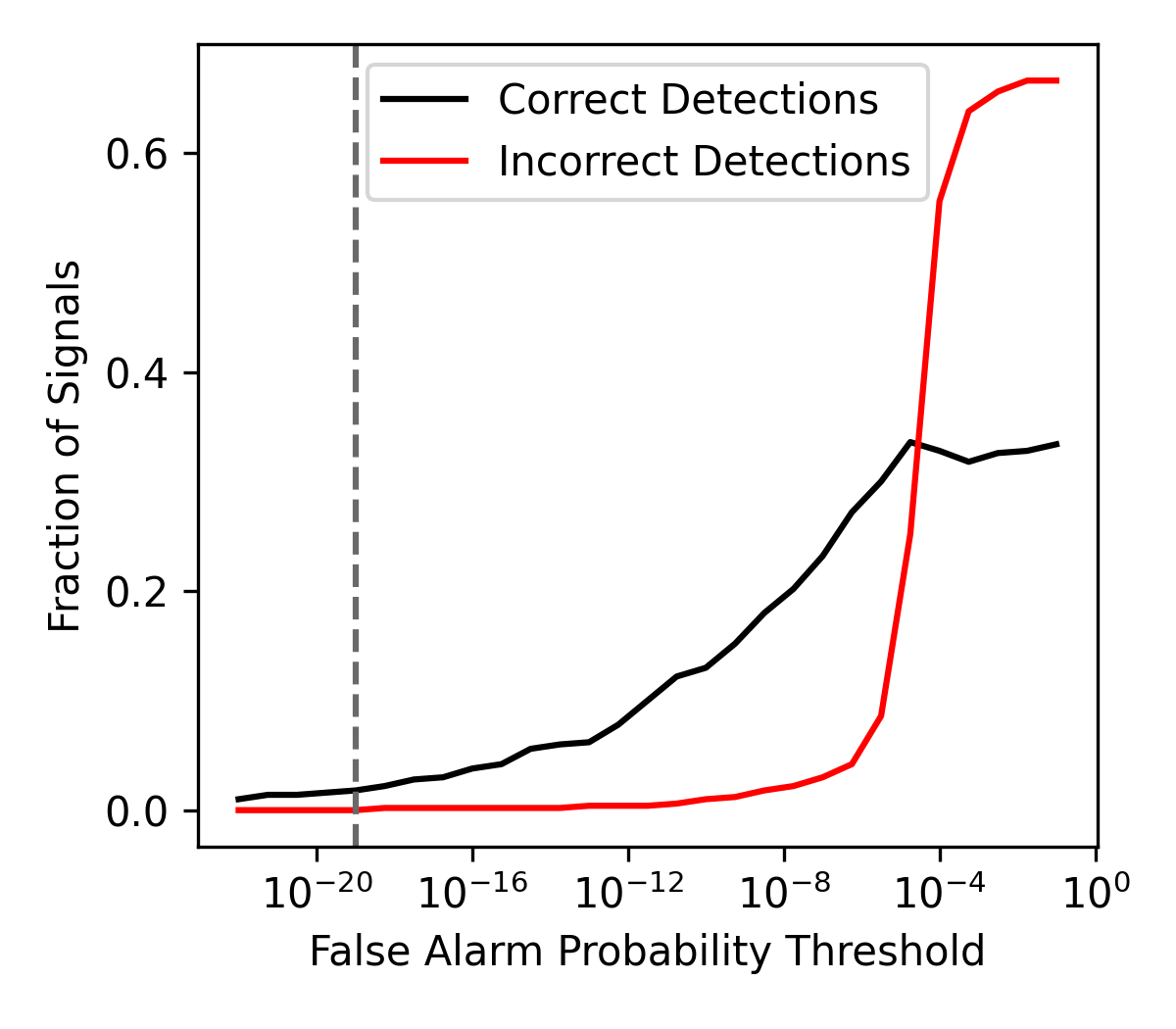}
    \caption{Empirical calibration of the FAP threshold for the traditional RV-level detection method. The black curve shows the fraction of injected planets correctly recovered, while the red curve shows the fraction of trials in which the accepted candidate does not match the injected signal. The dashed vertical line marks the adopted threshold, ${\rm FAP} < 10^{-19}$, chosen as the largest value that yields zero spurious detections in the injection--recovery sample.
    }
    \label{fig:fap_calibration}
\end{figure}

\begin{figure*}[htbp]
    \centering
    \includegraphics[width=\textwidth,trim={0cm 0cm 0cm 0.0cm}, clip]{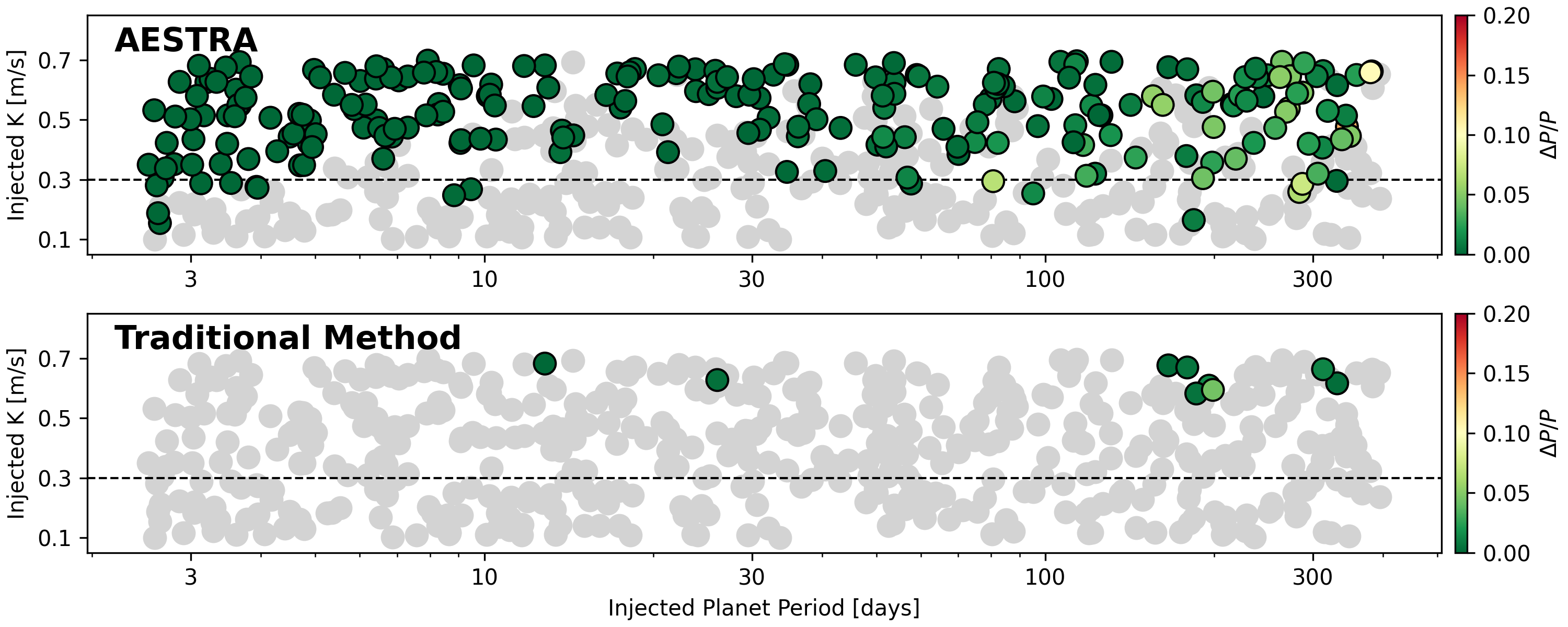}
    \caption{
    Detection performance in period–amplitude space for the injection--recovery sample. Each point represents one injected planetary signal. Colored points denote successful detections, with color indicating the fractional period error, and gray points indicate non-detections. Detection thresholds for both methods are calibrated to yield zero spurious detections in this injection--recovery suite.
     \textit{Top:} Recovery performance of injected planets using radial velocities corrected with the \aestra model, which recovers 238 out of 500 injected signals across a broad range of periods and amplitudes. 
     \textit{Bottom:} Recovery performance for injected planets using CCF-derived radial velocities corrected with traditional stellar activity indicators. The method recovers only 9 signals under the same false-positive constraint. The horizontal dashed line marks $K = 0.3$ \metersec.
    }
    \label{fig:injection_recovery}
\end{figure*}

\subsubsection{FAP Threshold for the Traditional Method}
\label{subsubsec:calibrate_fap}

We also evaluate the traditional RV-level search described in \autoref{subsec:traditional_candidate_search} as a standalone detection method. For this search, candidate detections are selected from a generalized Lomb--Scargle periodogram using a false-alarm probability (FAP) threshold.

Before computing the periodogram, we apply daily binning to the detrended RVs. This step is standard in RV-level analyses, where multiple exposures within a night are often averaged to reduce short-timescale stellar variability, including acoustic oscillations and granulation \citep[e.g.,][]{2011A&A...525A.140D, 2023MNRAS.525.1687J}. It is especially important here because the NEID solar time series is densely sampled within each observing day. If the periodogram is computed directly from the exposure-level RVs, the large number of closely spaced measurements and their correlated short-timescale structure can produce extremely small formal FAP values for peaks that are not necessarily physically significant. Daily binning reduces the leverage of this intraday sampling and yields a better-behaved RV time series for periodogram-based detection.

We compute generalized Lomb--Scargle periodograms of the daily binned, detrended RVs using the \texttt{astropy} implementation \citepalias{2018AJ....156..123A}, and assign a FAP to each candidate periodic signal. To place the traditional method and \aestra on comparable footing, we calibrate the FAP threshold empirically using the injection--recovery sample. \autoref{fig:fap_calibration} shows the fraction of correctly recovered injections and the fraction of spurious detections as a function of the adopted FAP threshold.

As the FAP threshold is relaxed, the recovery rate increases, but the spurious detection rate also rises. We adopt the largest FAP threshold that yields zero spurious detections among the 500 injection trials, as indicated by the dashed vertical line in \autoref{fig:fap_calibration}, to match the criteria for the \aestra injection-recovery test. This choice corresponds to
${\rm FAP} < 10^{-19}$.

\subsection{Success Criteria}
\label{subsec:success_metric}
For each injection case, the detection pipeline may return zero, one, or multiple candidate periodic signals. For the purpose of evaluating the single-planet injection–recovery performance, we reduce this list to a single candidate before comparing with the injected signal. Specifically, we rank the candidates that pass the method-specific detection threshold by their detection significance (i.e. $Q_{\rm detect}$ or FAP value) and retain only the top-ranked candidate. Lower-ranked candidates from the same trial are not used in the recovery accounting. If no candidate passes the detection threshold, the trial is classified as a non-detection.

The retained candidate is considered a successful recovery if its recovered period agrees with the injected period to within 20\%, and its recovered semi-amplitude is broadly consistent with the injected value, requiring $|K_{\rm rec}-K_{\rm inj}|/K_{\rm inj}<1$. 
Accepted candidates that do not match the injected signal are classified as spurious detections, or false positives. We do not impose an additional phase-coherence requirement in the primary recovery criterion because the inferred phase is strongly covariant with the recovered period: over the multi-year observing baseline, even a small period offset can produce a large phase difference.

The amplitude requirement should be interpreted as a loose consistency check. In the present \aestra training, multiple initialized sinusoidal candidates compete with the activity model and with one another. This procedure is designed to identify the period of a coherent Doppler signal that survives the activity decomposition, not to provide a final orbital solution at a known period. This competition can redistribute signal power among candidate components and bias individual fitted amplitudes low. Conversely, in the traditional RV-level search, the raw RV scatter is at the $\sim 2~{\rm m~s^{-1}}$ level, substantially larger than the injected sub-meter-per-second signals, so the best-fit amplitudes of weak candidates can be strongly biased high. For these reasons, we avoid a tight amplitude-recovery threshold, since our goal is to measure detection completeness rather than precise mass recovery.

Nevertheless, the successful \aestra recoveries generally have much more accurate amplitudes than the formal acceptance window allows: the 50th, 84th, and 95th percentiles of $|K_{\rm rec}-K_{\rm inj}|/K_{\rm inj}$ are 13.6\%, 27.1\%, and 43.1\%, respectively. As an additional check, we verified that the successful \aestra recoveries are phase-consistent with the injected signals: when the recovered and injected signals are compared at a common reference epoch, $t_{\rm ref}=200\,{\rm d}$, all successful recoveries agree to within 0.25 orbital cycles.


\subsection{Injection--Recovery Performance}
\label{subsec:injection_performabce}

\autoref{fig:injection_recovery} compares the detection performance of \aestra and a traditional periodogram-based approach across the injection--recovery sample. Each point represents one injected signal in period–amplitude space, and is colored according to whether the signal is successfully recovered. Detection criteria for both methods are calibrated to yield zero spurious detections in the 500 injection cases. 

\aestra recovers 238 out of 500 injected signals, with detections spanning a broad range of periods from a few days to several hundred days. In contrast, the traditional method recovers only 9 signals under the same false-positive constraint. At zero false positives, \aestra therefore achieves an order of magnitude increase in detection completeness. A striking difference between the two methods emerges at low signal amplitudes. Under the zero false-positive criterion, the traditional approach fails to recover any injected signals with $K < 0.5\,$\metersec, effectively defining a sensitivity floor at this level. In contrast, \aestra successfully recovers 13 signals in the range $K = 0.2\text{--}0.3\,$\metersec.

\begin{figure}[htbp]
    \centering
    \hspace*{-0.4cm}\includegraphics[width=0.47\textwidth,trim={0.1cm 0.1cm 0.0cm 0.0cm}, clip]{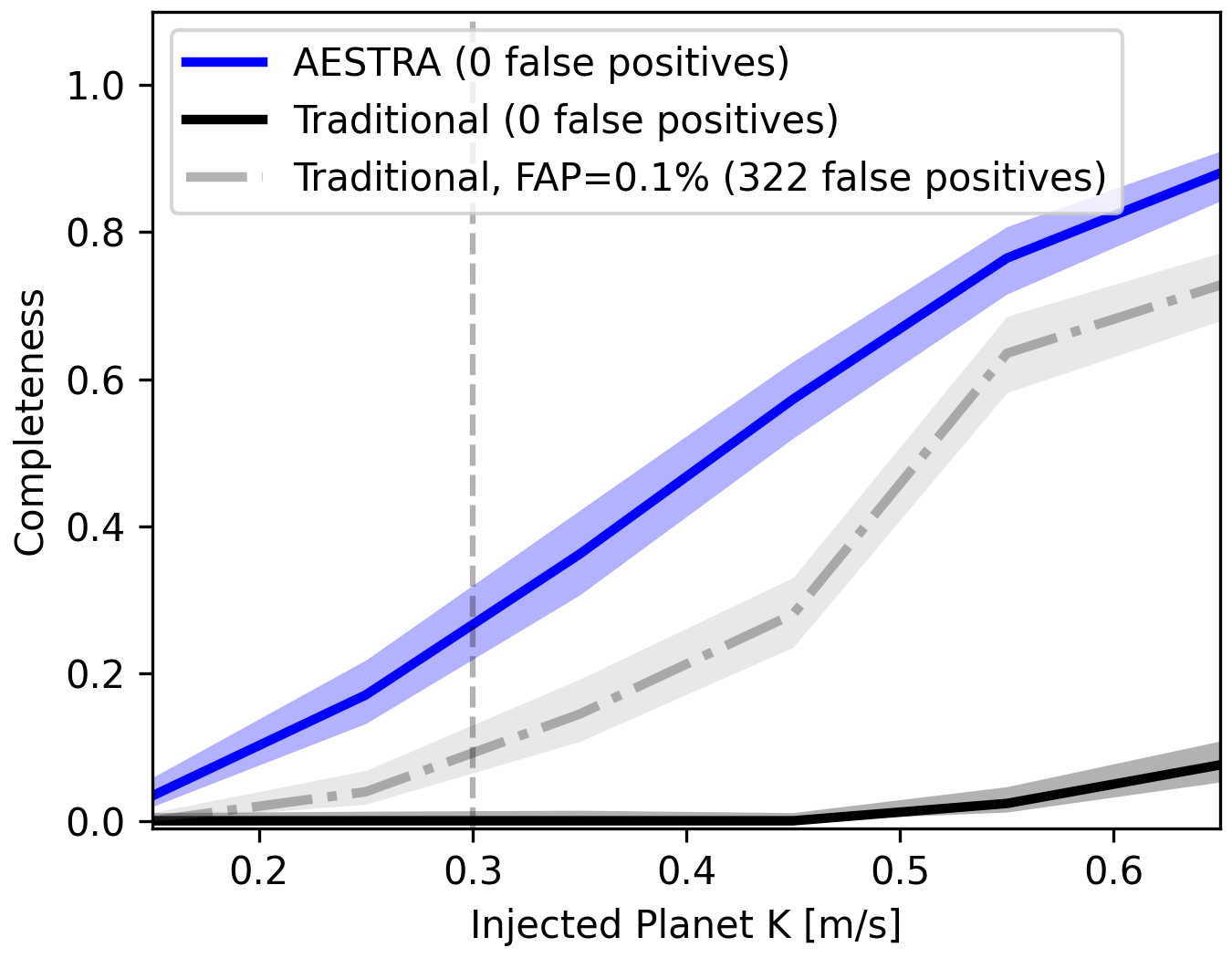}
    \caption{
    Detection completeness as a function of injected semi-amplitude $K$ in the injection--recovery experiments. The blue curve shows \aestra with the calibrated detection criterion, the black curve shows the traditional periodogram-based method calibrated to zero false positives, and the gray dashed curve shows the traditional method evaluated at ${\rm FAP}=0.1\%$. Shaded regions indicate Wilson binomial confidence intervals, and the vertical dashed line marks $K=0.3$ \metersec.
    }
    \label{fig:completeness_vs_K}
\end{figure}

\autoref{fig:completeness_vs_K} quantifies the detection performance as a function of injected semi-amplitude. \aestra achieves substantially higher completeness across the full amplitude range, with the largest gains at low $K$. In particular, for $K \lesssim 0.3$~\metersec, where stellar activity dominates the apparent RV variability, the traditional method recovers no signals under the zero false-positive constraint, while \aestra  already attains significant completeness. 
These detections span a wide range of orbital periods, from a few days to several hundred days, indicating that the gain in sensitivity is not confined to a specific timescale but reflects a systematic improvement across the parameter space.
If the traditional method is instead evaluated at the commonly used FAP $=0.1\%$ threshold, it recovers 159 injected signals but produces 322 spurious detections, as shown in \autoref{fig:fap_calibration}.

\section{Discussion}
\label{sec:discussion}

The central result of this work is that \aestra can distinguish low-amplitude coherent Doppler signals from activity-driven, wavelength-dependent spectral variability in real Sun-as-a-star spectra. 
Under a detection criterion calibrated to yield no false positives in the injection ensemble of 500 planets, \aestra recovers substantially more injected planets than the traditional approach, with the largest gains appearing near and below the $K=0.3\ $\metersec\ regime. 
The result therefore supports the premise that reaching Earth-analog sensitivity may be enabled by methods that exploit the full spectral information content of high-precision observations.

Several caveats are worth noting. First, the validation presented here uses the NEID solar feed, which provides an unusually favorable data set: high cadence, high signal-to-noise ratio, dense temporal sampling, and a single well-characterized star. The performance of \aestra on fainter stars, more active stars, rapidly rotating stars, or sparsely sampled surveys remains to be tested. Second, the detection boundary used here is empirically calibrated through injection--recovery tests. This is a strength for controlled validation, but it also means that practical applications will require system-specific calibration of completeness and false-positive behavior.

In this work, the injected and recovered signals are assumed to be of a single planet on a circular orbit, and the planet search is initialized from candidate periods identified by the traditional RV-level search. This controlled setup isolates the activity--Doppler separation problem but does not test the additional degeneracies introduced by eccentric or multi-planet systems. As a result, the present implementation cannot recover multiple planets simultaneously, nor can it recover a signal whose period is entirely missed by the initial candidate-generation stage.

A natural extension to multiple-planet systems would be to apply the search iteratively: high-amplitude planetary signals identified in earlier runs could be included in the model or subtracted before searching for lower-amplitude signals. More generally, the candidate-generation stage could be improved with a more complete period search or with neural network RV estimators trained directly on spectra, as implemented by \cite{2024AJ....167...23L}. These extensions are conceptually straightforward, but their completeness, false-positive behavior, and failure modes would need to be quantified with dedicated injection--recovery analyses. We leave such tests, together with eccentric-orbit recovery, to future work.

A more fundamental limitation is the degeneracy between stellar activity and planetary signals when both produce coherent variability at similar periods. This limitation is not unique to \aestra, nor to machine-learning methods. If a long-lived active region, rotational harmonic, or magnetic cycle produces an RV signal close to the period of a planet candidate, the data may not contain enough information to define a unique decomposition. In such cases, additional temporal coverage, independent activity diagnostics, or repeated observing seasons may be required to determine whether the candidate remains phase-coherent while the activity signal evolves. This failure mode motivates conservative detection criteria and reinforces the importance of properly treating ambiguous recoveries and non-detections as informative outcomes.

Finally, the present version of \aestra II is focused on validating the spectral disentanglement and activity-inference components. Unlike the original \aestra framework described in \cite{2024AJ....167...23L}, it does not train a neural-network RV estimator for computational scalability. 
A fully integrated framework that combines spectral disentanglement and neural RV estimation should further increase sensitivity at the expense of computational cost. We will pursue this direction in the future.

\section{Conclusion and Outlook}
\label{sec:conclusion}

We have presented \aestra II, a generative spectral-level framework for disentangling stellar activity, telluric absorption, continuum variability, and Doppler signals in extreme-precision radial-velocity observations. Applied to NEID Sun-as-a-star spectra, the model empirically removes telluric and continuum variability, preserves stellar line-shape information, and learns a compact latent representation of activity-driven spectral variability. This representation is then used to model the activity contribution to the measured RVs while separating candidate planetary signals.

In 500 single-planet injection--recovery experiments, \aestra substantially improves low-amplitude planet recovery under strict false-positive control. When the detection criterion is calibrated to produce zero spurious detections, \aestra recovers 238 injected planets, compared with 9 recovered by the traditional RV-level activity-correction baseline. Evaluating the traditional baseline at a fixed periodogram threshold of FAP $=0.1\%$ increases its apparent completeness, but produces 322 spurious detections in the same experiment. 
These results demonstrate the advantage of spectrum-level generative modeling: by exploiting spectral information that is lost in RV-level reductions, \aestra achieves substantially higher sensitivity while maintaining strict false-positive control.

The present study uses the Sun as a controlled benchmark target, with dense sampling, high signal-to-noise spectra, and well-defined injected signals. The next step is to determine how this performance extends to other stars, especially across different activity levels, rotation periods, spectral types, observing cadences, and signal-to-noise regimes. Another important direction is to incorporate explicit temporal structure into the learned activity representation. Spectrum-level models and time-domain models provide complementary information: the former learn how stellar activity changes the spectrum, while the latter learn how those changes evolve in time. Combining \aestra with Gaussian-process models, latent dynamical models, or other time-domain approaches may therefore improve robustness in regimes where activity and planetary signals are partially degenerate.

These developments are directly relevant for next-generation extreme-precision RV surveys targeting Earth-mass planets around Sun-like stars, especially the Terra Hunting Experiment \citep{thompson2016harps3} and the Second Earth Spectrograph \citep{2024SPIE13096E..8ES}. More broadly, spectrum-level generative modeling may improve the interpretation of marginal planet candidates, extend Doppler sensitivity to lower-mass planets around active stars, and provide a framework for causally separating complex spectral variability from multiple physical sources. By using the full information content of high-resolution spectra, \aestra provides a practical path toward detecting low-amplitude planetary signals that are currently buried beneath the stellar activity noise floor.

\begin{acknowledgments}
We thank Malena Rice, Earl Bellinger, and Tiger Lu for helpful discussions and feedback that improved this work.
This work made use of observations obtained with the NEID Solar Feed on the WIYN 3.5 m Telescope at Kitt Peak National Observatory, a facility of the NSF NOIRLab.
Computations for this work were performed using Princeton Research Computing resources at Princeton University.
We used ChatGPT, developed by OpenAI, for manuscript language editing and grammar checking \citepalias{ChatGPT2026}.
This project was supported by the Heising-Simons Foundation through grant 2025-5891.
\end{acknowledgments}
\newpage
\begin{contribution}

Yan Liang conceived the project, developed the methodology, performed the analysis, and wrote the manuscript. 
Joshua N.\ Winn co-conceived the project, provided supervision, and contributed to the interpretation of the results and the writing of the manuscript. 
Peter Melchior contributed to the conceptual development of the methodology and provided guidance on model design and analysis. 
Sicong Lu contributed to model development, including training and regularization strategies, and provided feedback on the methodology and results.
Quang Tran contributed to the design of the injection--recovery tests, the definition of detection criteria, and the interpretation of the recovery performance under different false-positive calibrations.


\end{contribution}

%
\facilities{WIYN (NEID)}


\software{
    \aestra \citep{2024AJ....167...23L}, 
    \texttt{spender} \citep{melchior2023autoencoding}, 
    Astropy \citepalias{2013A&A...558A..33A,2018AJ....156..123A,2022ApJ...935..167A}}


\newpage
\appendix

\section{Appendix information}
\subsection{Empirical Wavelength Corrections}
\label{appendix:wavelengths}

After the per-order formatting, masking, and normalization steps, each spectrum in a given echelle order is represented by its Earth-frame wavelength grid $\bm\lambda_{\mathrm{raw},i}$, normalized flux values $\mathbf{y}_{\mathrm{raw},i}$, and normalized flux weights $\mathbf{w}_{\mathrm{raw},i}$. The next step is to place all spectra on a common stellar-rest-frame wavelength grid. This approximately aligns the stellar absorption lines across epochs by removing the dominant relative motion between the Earth and the Sun. For this purpose, the correction does not need to recover the exact physical stellar rest frame at each epoch; it only needs to reduce apparent line shifts from the ${\rm km\,s^{-1}}$ scale to the \metersec\ scale, so that the remaining variability can be modeled spectrally. For real exoplanet hosts, this step would remove only the Earth's barycentric motion, leaving any stellar reflex motion in the data.

A practical complication is that spectra obtained before and after the 2022 Kitt Peak wildfire have different wavelength solutions. To enable joint modeling within a single framework, we introduce an empirical wavelength correction for the pre-fire data before shifting all spectra onto the common stellar-rest-frame grid. For each order, we define a smooth cubic-spline function, $f_{\mathrm{spline}}$, with seven trainable coefficients shared across all pre-fire spectra in that order. For an individual spectrum $i$, the resulting fractional wavelength correction is
\begin{equation}
\bm\delta_{\mathrm{spline},i}
=
f_{\mathrm{spline}}(\bm\lambda_{\mathrm{raw},i}),
\end{equation}
where $\bm\delta_{\mathrm{spline},i}$ is the spline evaluated on the raw wavelength grid of spectrum $i$.

We then construct the shifted wavelength grid as
\begin{equation}
\bm\lambda_{\mathrm{shifted},i} =
\begin{cases}
\bm\lambda_{\mathrm{raw},i}\left(1 + v_{\mathrm{ssb},i}/c\right), & \text{post-fire}\\[4pt]
\bm\lambda_{\mathrm{raw},i}\odot\left(1 + v_{\mathrm{ssb},i}/c + \bm\delta_{\mathrm{spline},i}\right), & \text{pre-fire}
\end{cases}
\end{equation}
where $v_{\mathrm{ssb},i}$ is the NEID barycentric correction and $\odot$ denotes element-wise multiplication. This empirical spline removes the dominant low-order mismatch between the pre-fire and post-fire wavelength solutions while leaving higher-order spectral variability to be captured by the downstream model.

Finally, each shifted spectrum is linearly interpolated onto a fixed stellar-rest-frame wavelength grid, $\bm\lambda_{\star}$:
\begin{align}
\mathbf{y}_{\mathrm{obs},i} &= \mathrm{Interp}\!\left(\bm\lambda_{\star};\, \bm\lambda_{\mathrm{shifted},i},\, \mathbf{y}_{\mathrm{raw},i}\right),\\
\mathbf{w}_{\mathrm{obs},i} &= \mathrm{Interp}\!\left(\bm\lambda_{\star};\, \bm\lambda_{\mathrm{shifted},i},\, \mathbf{w}_{\mathrm{raw},i}\right).
\end{align}

This produces a set of spectra and inverse-variance weights,
$(\mathbf{y}_{\mathrm{obs},i},\mathbf{w}_{\mathrm{obs},i})$, defined on the common grid. These quantities are the inputs to the order-level generative decomposition described in \autoref{subsec:joint}.

\subsection{Loss Functions}
\label{appendix:loss}

This section describes the loss function used for the order-level spectral decomposition introduced in \autoref{sec:decompose}. For each retained echelle order, the model reconstructs the observed spectrum as the product of three learned components: a telluric absorption component, a smooth continuum component, and a stellar line-shape variability component, together with a scalar additive offset. The goal of this step is to remove telluric and continuum contamination while preserving stellar line-shape variability in the corrected spectra.

For each order, we optimize the model parameters by minimizing
\begin{equation}
\mathcal{L}
=
\mathcal{L}_{\rm fid}
+
\mathcal{L}_{\rm reg},
\end{equation}
where $\mathcal{L}_{\rm fid}$ enforces accurate reconstruction of the observed spectra and $\mathcal{L}_{\rm reg}$ imposes weak morphology-based regularization on the three component branches.

The fidelity loss is the weighted reconstruction error between the observed spectrum $\mathbf{y}_{{\rm obs},i}$ and the model spectrum $\mathbf{y}_{{\rm model},i}$,

\begin{equation}
\label{eq:fid_loss}
\mathcal{L}_{\rm fid}=\sum_i
\frac{\sum_j w_{ij}\left(y_{{\rm obs},ij}-y_{{\rm model},ij}\right)^2}{\sum_j m_{ij}
},
\end{equation}
where $i$ indexes spectra within a batch and $j$ indexes wavelength bins. The binary mask $m_{ij}$ identifies valid pixels, with $m_{ij}=1$ when $w_{ij}>1$ and $m_{ij}=0$ otherwise. 
\textcolor{black}{The denominator normalizes the $\chi^2$-like reconstruction error by the number of valid pixels in each spectrum, so that spectra with different masked regions contribute on a comparable per-pixel basis.}

The regularization term is included to reduce degeneracies among the three branches. Without additional guidance, reconstruction alone would not uniquely determine whether a given residual structure should be assigned to the telluric, continuum, or stellar component. We therefore impose weak component-specific priors that encourage each branch to adopt the morphology expected from its physical origin. These terms do not use external telluric templates or synthetic stellar spectra; they only guide the decomposition toward sparse telluric absorption, smooth continuum variation, and localized stellar line-shape variability.

The telluric component is modeled as a non-negative absorption depth
\(\mathbf{d}_{t,i}\), with a single learned eigenspectrum whose amplitude
varies from exposure to exposure. This component is convolved with a
learnable LSF-like broadening kernel, $\mathbf{k}_t$, before being combined with the other
spectral components:
\begin{equation}
    \mathbf{y}_{t,i}={\rm Convolve}\left(\mathbf{d}_{t,i}, \mathbf{k}_t\right).
\end{equation}

We regularize the telluric component using
\begin{equation}
    \mathcal{L}_{t}=\sum_{i}\left[\min \left(\mathbf d_{t,i},\beta_t\right)\right]^2,
\end{equation}
where \(\beta_t\) is a tunable telluric-depth scale. This penalty behaves like an \(L_2\) penalty for weak absorption (\(\mathbf d_{t,i}<\beta_t\)) and saturates for stronger absorption (\(\mathbf d_{t,i}\geq\beta_t\)). It therefore suppresses low-amplitude continuum structure in the telluric component, while allowing localized absorption features to be represented when doing so substantially improves the spectral reconstruction. In combination with the low-dimensional telluric parameterization and the learned broadening convolution, this capped penalty encourages the telluric component to represent sparse, line-like absorption rather than broadband continuum variations. The parameter \(\beta_t\) controls the sensitivity of the model to weak telluric absorption. Smaller values allow the telluric component to activate for shallower micro-telluric lines, while larger values restrict the branch to stronger telluric features. The appropriate value depends on the signal-to-noise ratio and telluric contamination level of the data. For the NEID solar spectra analyzed here, we adopt \(\beta_t=0.05\).


Continuum variability is modeled using five learned eigenspectra together with a constant term. To restrict this branch to broad continuum-like structure, the continuum output is convolved with a fixed Gaussian smoothing kernel $\mathbf{k}_c$ with a characteristic scale of approximately 200 pixels, comparable to the fringe-like residuals seen in the blaze-corrected spectra. 
Let $\mathbf{\tilde y}_{c,i}$ denote the unconvolved continuum output for exposure $i$. The continuum component is
\begin{equation}
\mathbf{y}_{c,i}={\rm Convolve}\left(\mathbf{\tilde y}_{c,i}, \mathbf{k}_c\right),
\end{equation}
where $\mathbf{k}_c$ is the broad smoothing kernel.
This smoothing plays a role similar to imposing a fixed correlation length in wavelength, analogous in spirit to a Gaussian Process smoothness prior. However, the model is not nearly as flexible as a Gaussian Process: the continuum is represented by a small number of learned basis spectra with exposure-dependent amplitudes, rather than by a flexible stochastic function drawn from a covariance kernel. 

The five learned eigenspectra and the smoothing kernel are intended to capture epoch-to-epoch variations in the blaze-corrected spectra that are too broad in wavelength to be plausibly stellar. 
We regularize the continuum component by requiring its mean over the training batch to remain close to zero,
\begin{equation}
\mathcal{L}_{c}=N_{\rm spec}\sum_j\left[\bar y_{c,j}\right]^2 ,
\end{equation}
where $\bar y_{c,j}$ denotes the $j$-th wavelength bin of the batch-mean continuum spectrum. This penalty encourages static or nearly static offsets to be absorbed by the scalar term $b_i$ or by the mean stellar template, while allowing the continuum branch to describe time-variable broad-wavelength structure. Because the continuum branch is more flexible than the telluric branch, it receives a stronger penalty, as described below.

In the order-level decomposition stage, stellar variability is modeled using three eigenspectra per order. This component is included to capture and protect possible activity-related line-shape variability during the telluric and continuum correction steps, so that astrophysical spectral structure is not mistakenly assigned to those nuisance components. To encourage the stellar component to represent localized line-profile distortions rather than broadband flux variations, we high-pass filter  the stellar component by subtracting a version smoothed with a Gaussian kernel of width $\sigma=25$ pixels. We further regularize it according to
\begin{equation}
\mathcal{L}_{\star}=\sum_{i,j}m^{\rm cont}_{ij}\left(\Delta y_{\star,ij}\right)^2 ,
\end{equation}
where $m^{\rm cont}_{ij}=1$ when $y_{{\rm obs},ij}/(1-y_{t,ij})>0.9$ and $m^{\rm cont}_{ij}=0$ otherwise. This term penalizes the stellar component in wavelength bins identified as being near the continuum after telluric correction, encouraging the stellar component to concentrate its power near deep stellar absorption lines, where activity-driven line-profile distortions are expected to appear. Because the stellar branch is the most flexible component, it also requires the strongest regularization, as described below.

The total regularization loss is
\begin{equation}
\label{eq:total-loss}
\mathcal{L}_{\rm reg}=\alpha_{t} \mathcal{L}_{t}
+\alpha_{c}\mathcal{L}_{c}
+\alpha_{\star}\mathcal{L}_{\star}.
\end{equation}
In this work, we adopt $\alpha_t \ll \alpha_c \ll \alpha_\star$, with
$\alpha_t=0.02$, $\alpha_c=1$, and $\alpha_\star=10$. This hierarchy discourages the more flexible stellar branch from absorbing nuisance variability that can be explained by the simpler telluric or continuum branches. The specific values used here are tuned to the morphology and signal scales of the NEID solar data and would likely require adjustment for other instruments or stars. 

Finally, we emphasize that the stellar component inferred in this order-level decomposition is not used as the downstream stellar-activity representation for RV inference. Its role is only to protect stellar line-shape variability during the construction of activity-preserving spectra. The full-spectrum activity representation used in the RV analysis is learned later from the cleaned and merged spectra. In this context, conservative regularization of the order-level stellar branch is acceptable and helps stabilize the decomposition without removing the stellar variability needed for the subsequent activity model.

\subsection{CCF-based Diagnostics}
\label{appendix:ccf}
We compute cross-correlation functions (CCFs) from the cleaned and merged spectra to derive classical line-shape diagnostics. The CCFs are evaluated on a velocity grid from $-15\ {\rm km\,s^{-1}}$ to $+15\ {\rm km\,s^{-1}}$ using 201 uniformly spaced points.
From each normalized CCF, we extract four summary statistics: the line depth, Gaussian width (FWHM), bisector span (BIS SPAN), and an additive offset term. The depth, width, and offset are obtained from a Gaussian fit to the CCF profile. The BIS SPAN is defined as the velocity difference between the average bisector measured in the upper $10\%$--$40\%$ of the line depth and that in the lower $60\%$--$90\%$ of the line depth, following \citet{2014ApJ...796..132D}.

These CCF-based indicators provide a baseline representation of stellar activity for comparison with the spectral-level latent model, and are used for detrending in the traditional periodogram-based analysis described in the main text.

Although radial velocities can also be derived from the CCFs, we find that they exhibit slightly larger scatter than the template-fitting measurements adopted in this work. This is likely due to the use of the average spectrum as the correlation template, rather than a line mask optimized for high-precision RV extraction. For this reason, we adopt the template-fitting estimates as the fiducial apparent radial velocities throughout.

\subsection{Comparisons Between Stellar Latent Variables and Traditional Activity Indicators}
\label{appendix:latent}

\begin{figure}[htbp]
    \centering
    \includegraphics[width=\textwidth,trim={0cm 0.5cm 0cm 0cm}, clip]{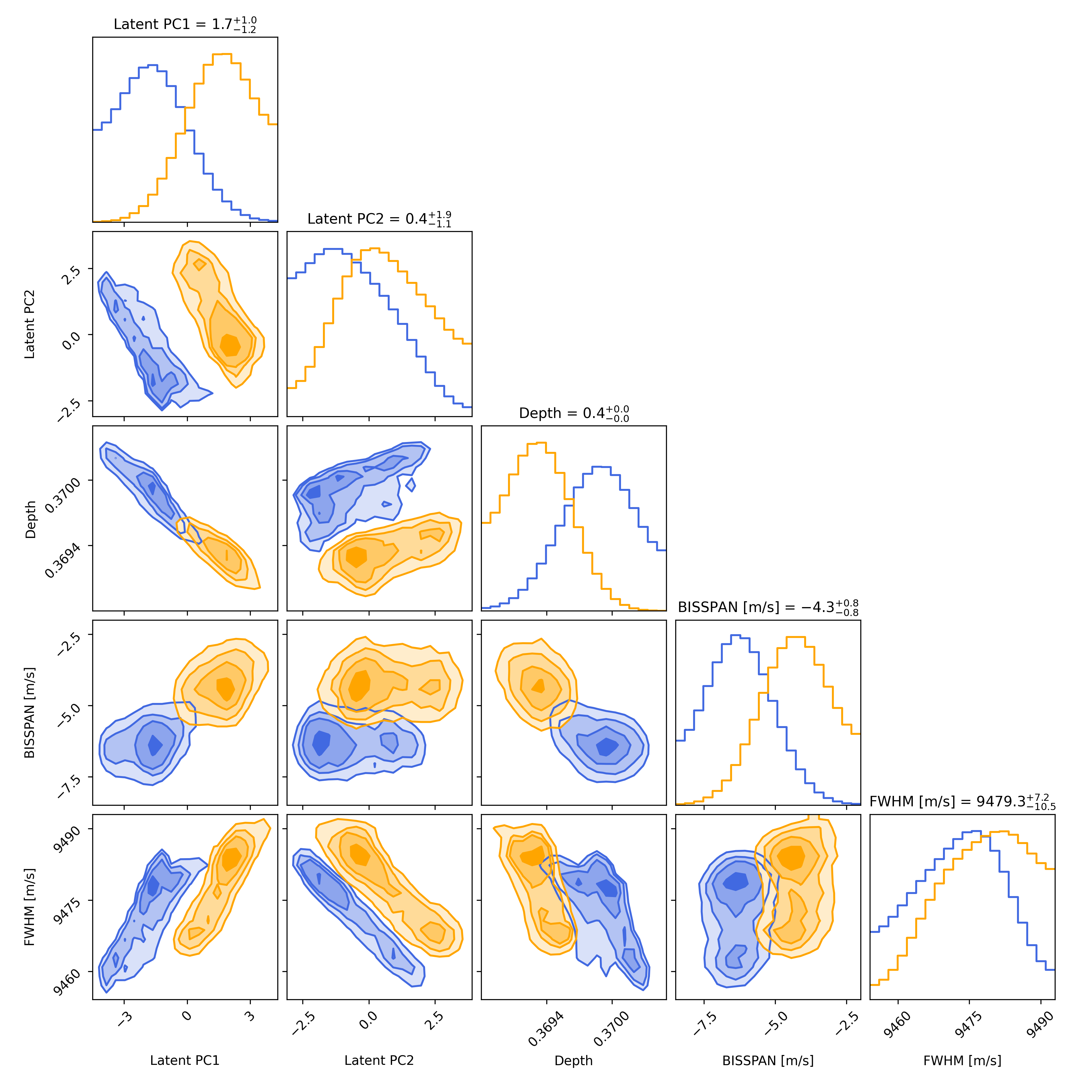}
    \caption{Corner plot comparing the first two principal components of the learned stellar latent representation with traditional CCF-based activity indicators, shown separately for the pre-fire (blue) and post-fire (orange) subsets. The two populations occupy partially distinct regions in both latent space and classical indicator space. In particular, the separation between the pre-fire and post-fire data is visible not only in the latent principal components, but also in combinations of traditional observables such as CCF depth and BIS SPAN. This behavior indicates that the unified latent representation preserves meaningful distinctions between observing regimes rather than artificially forcing the two subsets into a single homogeneous distribution. The different locations of the two populations are consistent with the known change in instrumental state after the 2022 fire, as well as with secular evolution in the dominant patterns of solar activity over the observing baseline.}
    \label{fig:latent_corner}
\end{figure}
\label{appendix_latent_corner}

For additional insight into the learned representation, \autoref{fig:latent_corner} compares the first two principal components of the stellar latent vectors with several classical CCF-based activity diagnostics, shown separately for the pre-fire and post-fire subsets. The two subsets occupy partially distinct regions of both the latent space and the traditional indicator space. This separation is not unexpected: the pre-fire and post-fire observations differ both in instrumental state and in the dominant patterns of solar variability sampled over time. In this sense, the latent representation is not simply memorizing a single activity axis, but is flexible enough to accommodate heterogeneous observing conditions within a unified model while still preserving meaningful structure in the data. The fact that a similar separation is also present in traditional indicators, particularly in the depth--BIS SPAN plane, further suggests that these differences reflect real changes in the underlying data rather than artifacts introduced by the latent model itself.

\subsection{Comprehensive Period Search Procedure}
\label{appendix:period_search}
To improve the completeness of the initial set of candidate periodic signals, we use a staged period-search procedure rather than a single periodogram of the full RV time series.

First, we split the dataset into pre- and post-fire subsets ($t<800 \, {\rm d}$ and $t>800\, {\rm d}$) and perform standard linear detrending against conventional activity indicators separately on each subset before computing generalized Lomb--Scargle periodograms. This step improves sensitivity by accounting for the change in instrument state across the wildfire shutdown and reducing activity-related variance within each segment.

Second, to enhance sensitivity to weak short-period signals that may be locally visible only during quieter intervals, we divide the apparent RV time series into four contiguous chunks separated at $t=500\, {\rm d}, 800\, {\rm d}$ and $1200\, {\rm d}$, and search each chunk independently for candidate periods with $P<10\,{\rm d}$.

Third, we perform additional searches for intermediate-period signals on the two broader subsets $t<800\, {\rm d}$ and $t>800\, {\rm d}$, retaining candidates with $P<150\,{\rm d}$. We also search the full time series for longer-period candidates in the range $90\, {\rm d} < P < 400\,{\rm d}$.

After collecting all candidate periods from these searches, we fit a sinusoid to the full detrended RV time series initialized at each candidate period. We then sort the resulting periods and merge near-duplicates, defining two neighboring solutions as distinct only if their fractional period difference exceeds 10\%. The remaining unique periods are used to initialize the sinusoidal components of the planetary RV module.


\bibliography{main}{}
\bibliographystyle{aasjournalv7}



\end{document}